\documentclass[ALICE,manyauthors]{cernphprep}

\usepackage[comma,square,numbers,sort&compress]{natbib}
\usepackage{hyperref}
\usepackage{lineno}
\usepackage{color}
\usepackage{floatrow}
\usepackage{verbatim}
\usepackage{amssymb}
\usepackage[T1]{fontenc}

\begin{document}

% $Id: commands.tex 934 2013-06-19 20:56:45Z mfloris $

\newcommand{\ITS}          {\mathrm{ITS}}
\newcommand{\TOF}          {\mathrm{TOF}}
\newcommand{\ZDC}          {\mathrm{ZDC}}
\newcommand{\ZDCs}         {\mathrm{ZDCs}}
\newcommand{\ZNA}          {\mathrm{ZNA}}
\newcommand{\ZNC}          {\mathrm{ZNC}}
\newcommand{\SPD}          {\mathrm{SPD}}
\newcommand{\SDD}          {\mathrm{SDD}}
\newcommand{\SSD}          {\mathrm{SSD}}
\newcommand{\TPC}          {\mathrm{TPC}}
\newcommand{\VZERO}        {\mathrm{V0}}
\newcommand{\VZEROA}       {\mathrm{V0-A}}
\newcommand{\VZEROC}       {\mathrm{V0-C}}
\newcommand{\pip}          {$\pi^{+}$}
\newcommand{\pim}          {$\pi^{-}$}
\newcommand{\kap}          {K$^{+}$}
\newcommand{\kam}          {K$^{-}$}
\newcommand{\he}{$^{3}{\mathrm{He}}$}
\newcommand{\ahe}{$^{3}\overline{\rm He}$}
\newcommand{\heFour}{$^{4}{\mathrm{He}}$}
\newcommand{\aheFour}{$^{4}\overline{\rm He}$}

\newcommand{\ad}{$\overline{\rm d}$}
\newcommand{\pbar}         {$\mathrm{\overline{p}}$}
\newcommand{\kzero}        {\ensuremath{\mathrm{ K^{0}_{S}}}}
\newcommand{\vzero}        {\ensuremath{\mathrm{V}^0}}
\newcommand{\lmb}          {\ensuremath{\Lambda}}
\newcommand{\almb}         {\ensuremath{\bar{\Lambda}}}
\newcommand{\allpart}      {$\pi^{\pm}$, K$^{\pm}$, \kzero, p(\pbar) and \lmb(\almb)}
\newcommand{\allpikp}      {$\pi^{\pm}$, K$^{\pm}$ and p(\pbar)}
\newcommand{\pikp}         {$\pi$, K, and p}
\newcommand{\allpi}        {$\pi^{\pm}$}
\newcommand{\allk}         {K$^{\pm}$}
\newcommand{\allp}         {p(\pbar)}
\newcommand{\alllmb}       {\lmb(\almb)}
\newcommand{\degree}       {$^{\mathrm{o}}$}
\newcommand{\dg}           {\mbox{$^\circ$}}
\newcommand{\dedx}         {\ensuremath{\mathrm{d}E/\mathrm{d}x}}
\newcommand{\dndy}         {d$N$/d$y$}
\newcommand{\pp}           {pp}
\newcommand{\ppbar}        {\mbox{$\mathrm {p\overline{p}}$}}
\newcommand{\PbPb}         {\mbox{Pb--Pb}}
\newcommand{\pPb}          {\mbox{p--Pb}}
\newcommand{\AuAu}         {\mbox{Au--Au}}
\newcommand{\pseudorap}    {\mbox{$\left | \eta \right | $}}
\newcommand{\dNdeta}       {\ensuremath{\mathrm{d}N_\mathrm{ch}/\mathrm{d}\eta}}
\newcommand{\dNdy}         {\ensuremath{\mathrm{d}N_\mathrm{ch}/\mathrm{d}y}}
\newcommand{\dNdyst}       {\ensuremath{\sqrt{\frac{dN_\pi/dy}{s_T}}}}
\newcommand{\dNdetatr}     {\mathrm{d}N_\mathrm{tracklets}/\mathrm{d}\eta}
\newcommand{\dNdetar}[1]   {\mathrm{d}N_\mathrm{ch}/\mathrm{d}\eta\left.\right|_{|\eta|<#1}}
\newcommand{\dNdetamean}   {\ensuremath{\langle\dNdeta  \rangle}}
\newcommand{\lum}          {\, \mbox{${\mathrm{ cm}}^{-2} {\mathrm {s}}^{-1}$}}
\newcommand{\barn}         {\, \mbox{${\mathrm{barn}}$}}
\newcommand{\m}            {\, \mbox{${\mathrm{m}}$}}
\newcommand{\ncls}         {\ensuremath{N_{cls}}}
\newcommand{\nsigma}       {\ensuremath{n\sigma}}
\newcommand{\dcaxy}        {\ensuremath{\mathrm{DCA_{\rm xy}}}}
\newcommand{\dcaz}         {\ensuremath{\mathrm{DCA_{\rm z}}}}
\newcommand{\EcrossB}      {E$\times$B}%{\ensuremath{{\rm E}\times{\rm B}}}
\newcommand{\bb}           {Bethe-Bloch}
\newcommand{\s}            {\ensuremath{\sqrt{s}}}
\newcommand{\PT}           {\ensuremath{p_{\mathrm{T}}}}
\newcommand{\MT}           {\ensuremath{m_{\mathrm{T}}}}
\newcommand{\hlab}         {\ensuremath{\eta_{\mathrm{lab}}}}
\newcommand{\ynn}         {\ensuremath{y_{\mathrm{NN}}}}
\newcommand{\ycms}         {\ensuremath{y_{\mathrm{CMS}}}}
\newcommand{\ylab}         {\ensuremath{y_{\mathrm{lab}}}}
\newcommand{\ppi}          {\ensuremath{{\mathrm{p}}/\pi}}
\newcommand{\kpi}          {\ensuremath{{\mathrm{K}}/\pi}}
\newcommand{\lpi}          {\ensuremath{{\mathrm{ \Lambda}}/\pi}}
\newcommand{\lks}          {\ensuremath{{\mathrm{ \Lambda}}/\mathrm{{K}^{0}_{S}}}}
\newcommand{\mt}           {\ensuremath{m_{\mathrm{T}}}}
\newcommand{\snn}          {\ensuremath{\sqrt{s_{\mathrm{NN}}}}}
\newcommand{\snnbf}        {\ensuremath{\mathbf{{\sqrt{s_{\mathbf{ NN}}}}}}}
\newcommand{\sonly}        {\ensuremath{\sqrt{s}}}
\newcommand{\Npart}        {\ensuremath{N_\mathrm{part}}}
\newcommand{\avNpart}      {\ensuremath{\langle N_\mathrm{part} \rangle}}
\newcommand{\avNpartdata}  {\ensuremath{\langle N_\mathrm{part}^{\mathrm{data}} \rangle}}
\newcommand{\Ncoll}        {\ensuremath{N_\mathrm{coll}}}
\newcommand{\avNcoll}      {\ensuremath{\langle N_\mathrm{coll} \rangle}}
\newcommand{\Dnpart}       {\ensuremath{D\left(\Npart\right)}}
\newcommand{\DnpartExp}    {\ensuremath{D_{\mathrm{exp}}\left(\Npart\right)}}
\newcommand{\dNdetapt}     {\ensuremath{\dNdeta\,/\left(0.5\Npart\right)}}
\newcommand{\dNdetaptr}[1] {\ensuremath{\dNdetar{#1}\,/\left(0.5\Npart\right)}}
\newcommand{\dNdetape}     {\left(\ensuremath{\dNdeta\right)/\left(\avNpart/2\right)}}
\newcommand{\dNdetaper}[1] {\ensuremath{\dNdetar{#1}\,/\left(\avNpart/2\right)}}
\newcommand{\dndydpt}      {\ensuremath{{\mathrm{d}}^2N/({\mathrm{d}}y {\mathrm{d}}p_{\mathrm{t}})}}
\newcommand{\abs}[1]       {\ensuremath{\left|#1\right|}}
\newcommand{\signn}        {\ensuremath{\sigma^{\mathrm{inel.}}_{\mathrm{NN}}}}
\newcommand{\vz}           {\ensuremath{V_{z}}}
\newcommand{\Tfo}          {\ensuremath{{T}_{\mathrm{kin}}}}
\newcommand{\Tch}          {\ensuremath{{T}_{\mathrm{ch}}}}
\newcommand{\bT}           {\ensuremath{\beta_{\mathrm{T}}}}
\newcommand{\avbT}         {\ensuremath{\langle \beta_{\mathrm{T}}\rangle}}
\newcommand{\avpT}         {\ensuremath{\langle \PT \rangle}}
\newcommand{\muB}          {\ensuremath{\mu_{B}}}
\newcommand{\stat}         {({\it stat.})}
\newcommand{\syst}         {({\it sys.})}
\newcommand{\Fig}[1]       {Fig.~\ref{#1}}
\newcommand{\green}[1]     {\textcolor{green}{#1}}
\newcommand{\blue}[1]      {\textcolor{blue}{#1}}
\newcommand{\red}[1]       {\textcolor{red}{#1}}
\newcommand{\white}[1]     {\textcolor{white}{#1}}
\newcommand{\gevc}         {\ensuremath{{\mathrm{GeV}}/c}}
\newcommand{\gevcsq}       {\ensuremath{{\mathrm{GeV}^4}/c^2}}
\newcommand{\mevc}         {\ensuremath{{\mathrm{MeV}}/c}}
\newcommand{\avg}[1]       {\ensuremath{\left\langle#1\right\rangle}}
\newcommand{\tkin}         {\ensuremath{T\mathrm{_{kin}}}}
\newcommand{\rppb}         {\ensuremath{R\mathrm{_{pPb}}}}
\newcommand{\rdau}         {\ensuremath{R\mathrm{_{dAu}}}}
\newcommand{\mtof}         {\ensuremath{{m_{\mathrm{TOF}}^2}}}
\newcommand{\dbard}        {\ensuremath{\mathrm{\bar{d}/d}}}
\newcommand{\pbarp}        {\ensuremath{\mathrm{\bar{p}/p}}}
\newcommand{\nbarn}        {\ensuremath{\mathrm{\bar{n}/n}}}
\newcommand{\alphas}       {\ensuremath{\alpha_{\mathrm{S}}}}
\newcommand{\mub}          {\ensuremath{\mathrm{\mu}_{\mathrm{B}}}}
\newcommand{\Btwo}         {\ensuremath{B_2}}
\newcommand{\geant}        {${\textsc{Geant4}}$\xspace}
\newcommand{\sigmainel}    {$\sigma_{\rm inel}(^{3}\overline{\rm{He}})$}

%\renewcommand{\labelitemi} {$-$}
%==========================================================%
%%% inline warnings for internal discussion 
%\newcommand{\warn}[1]      {\textbf{\textcolor{red}{[#1]}}}
\newcommand{\warn}[1]      {{\small\textbf{\textcolor{red}{(!\footnote{\textbf{(!)}~#1})}}}}
\newcommand{\warnin}[1]         {\textit{\textcolor{red}{(#1)}}}
%\newcommand{\warn}[1]      {#1}
%\newcommand{\warn}[1]      {{\small\textbf{(!\footnote{\textbf{(!)}~#1})}}\marginpar{\textbf{---}}}
%\newcommand{\todo}[1]      {\textbf{\textcolor{red}{[TODO: #1]}}}
%%% fake numbers
\newcommand{\fake}[1]      {\textbf{\textcolor{red}{#1}}}
\newcommand{\final}[1]     {\textbf{\textcolor{blue}{#1}}}
\newcommand{\prelim}[1]    {\textbf{\textcolor{magenta}{#1}}}
\renewcommand{\mod}[1]       {\textbf{\textcolor{red}{#1}}}

% to count the words
\newcommand{\detailtexcount}[1]{%
  \immediate\write18{texcount -merge -sum -q #1.tex output.bbl > #1.wcdetail }%
  \verbatiminput{#1.wcdetail}%
  \newpage
}
%%%%%%%%%%%%%%%  Title page %%%%%%%%%%%%%%%%%%%%%%%%
\begin{titlepage}
\PHyear{2022}
\PHnumber{023}      % required, will be obtained from PH
\PHdate{03 February}  % required, will be obtained from PH

\title{Measurement of \ahe\ nuclei absorption in matter \\ and impact on their propagation in the Galaxy}
\ShortTitle{Measurement of \ahe\ nuclei absorption in matter}

\Collaboration{ALICE Collaboration} %\thanks{See Appendix~\ref{app:collab} for the list of collaboration members}}
\ShortAuthor{ALICE Collaboration}

\begin{abstract}

In our Galaxy, light antinuclei composed of antiprotons and antineutrons can be produced through high-energy cosmic-ray collisions with the interstellar medium or could also originate from the annihilation of dark-matter particles that have not yet been discovered. On Earth, the only way to produce and study antinuclei with high precision is to create them at high-energy particle accelerators. Although the properties of elementary antiparticles have been studied in detail, the knowledge of the interaction of light antinuclei with matter is limited. We determine the disappearance probability of \ahe\ when it encounters matter particles and annihilates or disintegrates within the ALICE detector at the Large Hadron Collider. We extract the inelastic interaction cross section, which is then used as input to calculations of the transparency of our Galaxy to the propagation of \ahe\ stemming from dark-matter annihilation and cosmic-ray interactions within the interstellar medium. For a specific dark-matter profile, we estimate a transparency of about 50\%, whereas it varies with increasing \ahe\ momentum from 25\% to 90\% for cosmic-ray sources. The results indicate that \ahe\ nuclei can travel long distances in the Galaxy, and can be used to study cosmic-ray interactions and dark-matter annihilation.

\end{abstract}
\end{titlepage}
\setcounter{page}{2}

% \detailtexcount{main}

\section{Introduction}
There are no natural forms of antinuclei on Earth, but we know they exist because of fundamental symmetries in particle physics and their observation in interactions of high-energy accelerated beams.
Light antinuclei, objects composed of antiprotons ($\mathrm{\overline{p}}$) and antineutrons ($\mathrm{\overline{n}}$), such as $\mathrm{\overline{d}}$ ($\mathrm{\overline{p}\overline{n}}$), $^3\mathrm{\overline{\rm He}}$ ($\mathrm{\overline{p}\overline{p}\overline{n}}$)
and $^4\mathrm{\overline{\rm He}}$ ($\mathrm{\overline{p}\overline{p}\overline{n}\overline{n}}$), have been produced and studied at various accelerator facilities~\cite{SimonGillo:1995dh,Armstrong:2000gd,Afanasev:2000ku,Anticic:2004yj,Adler:2004uy,Alper:1973my,Henning:1977mt,Alexopoulos:2000jk,Aktas:2004pq,Asner:2006pw,Schael:2006fd,ALICE:2015wav,Adam:2015yta,Acharya:2017dmc,Acharya:2019rgc,Agakishiev:2011ib,Abelev:2010,STAR:2001pbk}, including precision measurements of the mass difference between nuclei and antinuclei~\cite{Adam:2015pna,Adam:2019phl}.
The interest in the properties of such objects is manifold. From the nuclear physics perspective, the production mechanism and interactions of antinuclei can elucidate the detailed features of the strong interaction that binds nucleons into nuclei~\cite{STAR:2015kha}. From the astrophysical standpoint, natural sources of antinuclei may include annihilation of dark-matter (DM) particles such as weakly interacting massive particles~\cite{Ibarra:2012cc} and other exotic sources such as antistars~\cite{Persic:1995ru,Poulin:2018wzu}.
DM constitutes about 27\% of the total energy density budget within our Universe~\cite{Aghanim:2018eyx}. This is demonstrated by the measurement of the fine structure of the cosmic microwave background~\cite{Bond:1984fp,deBernardis:2000sbo}, gravitational lensing of galaxy clusters~\cite{Clowe:2006eq} and the rotational curves of some galaxies~\cite{Persic:1995ru}.
Another possible source of antinuclei in our universe are high-energy cosmic-ray collisions with atoms in the interstellar medium.

The observation of antinuclei such as \ahe\ is one of the most promising signatures of DM annihilation of weakly interacting massive particles~\cite{Ibarra:2012cc,Carlson:2014ssa,Korsmeier:2017xzj,vonDoetinchem:2020vbj,Winkler:2020ltd}.
The kinetic-energy distribution of antinuclei produced in DM annihilation peaks at low kinetic energies ($E_{\rm kin}$ per nucleon $\lesssim 1$ GeV/$A$) for most assumptions of DM mass~\cite{Ibarra:2012cc}. In contrast, for antinuclei originating from cosmic-ray interactions the spectrum peaks at much larger $E_{\rm kin}$ per nucleon ($\simeq 10$ GeV/$A$). Thus, the low-energy region is almost free of background for DM searches.

To calculate the expected flux of antinuclei near Earth, one needs to know precisely the antinucleus formation and annihilation probabilities in the Galaxy.
The formation probability of light antinuclei (up to mass number $A = 4$) is currently studied at accelerators.
By now, several models successfully describe light-antinuclei production yields~\cite{Bellini:2020cbj,Kachelriess:2020amp,Braun-Munzinger:2018hat,Steinheimer:2012tb,Braun-Munzinger:1994zkz}. Such models are based on either the statistical-hadronization~\cite{ALICE:2015wav,Andronic:2010qu,Cleymans:2011pe,Vovchenko:2018fiy} or coalescence approach~\cite{Butler:1963pp,Scheibl:1998tk,Blum:2019suo,Mrowczynski:2016xqm,Mrowczynski:1994rn}.

Another crucial aspect in the search of antinuclei in our Galaxy is the knowledge of their disappearance probability when they encounter matter and annihilate or disintegrate. Antinuclei generated in our Galaxy may travel thousands of light years~\cite{Boschini:2020jty} before reaching the Earth and being detected. The journey of antinuclei through the Galaxy can be modelled by propagation codes, which incorporate the initial distribution of antinucleus sources, the interstellar gas distribution in the Galaxy, the elastic scatterings, and the inelastic hadronic interactions with the interstellar medium.
The antinucleus flux in the Solar System is further modulated by solar magnetic fields. During the entire journey, antinuclei can encounter matter and disappear. The disappearance probability is quantified through the inelastic cross section. It is normally studied employing particle beams of interest impinging on targets of known composition and thickness, but antinuclei beams are very challenging to obtain. Today, the Large Hadron Collider (LHC) is the best facility to study nuclear antimatter since its high energies allow one to produce, on average, as many nuclei as antinuclei in proton--proton (pp) and lead--lead (Pb--Pb) collisions~\cite{Abbas:2013rua,ALICE:2015wav}.
The detector material can serve as a target and the disappearance probability can be experimentally determined~\cite{Acharya:2020cee}.

This work presents the measurement of the \ahe\ inelastic cross section \sigmainel, obtained using data from the ALICE experiment. These results are used in model calculations to assess the effect of the disappearance of antinuclei during their propagation through our Galaxy. The associated uncertainties are estimated based on experimental data. The transparency of our Galaxy to the propagation of \ahe\ nuclei stemming from a specific DM source and from interactions of high-energy cosmic rays with the interstellar medium is determined, providing one of the necessary constraints for the study of antinuclei in space.

\section{Determination of the inelastic cross section}

\begin{figure}[htbp]
\centering
\includegraphics[width=0.9\textwidth]{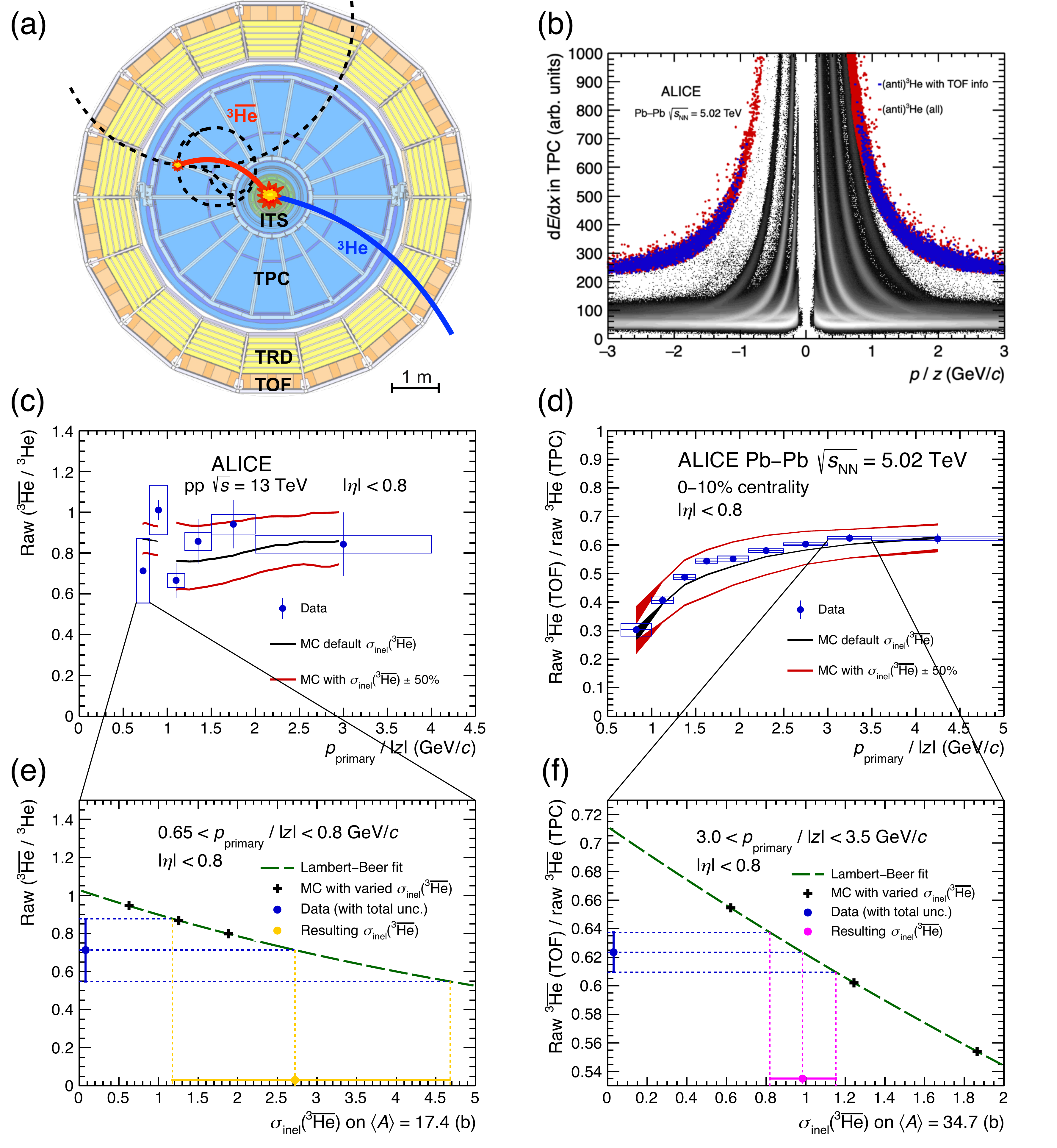}
\caption{Description of the steps followed for the extraction of \sigmainel. \textbf{(a)} Schematic of the ALICE detectors at midrapidity in the plane perpendicular to the beam axis, with the collision point located in the middle; the ITS, TPC, TRD and TOF detectors are shown in green, blue, yellow and orange, respectively. A \ahe\ which annihilates in the TPC gas is shown in red, and a \he\ that does not undergo an inelastic reaction and reaches the TOF detector in blue; the dashed curves represent charged (anti)particles produced in the \ahe\ annihilation. \textbf{(b)} Identification of (anti)nuclei by means of their specific energy loss d$E$/d$x$ and momentum measurement in the TPC. The red points show all (anti)$^{3}$He nuclei reconstructed with the TPC detector, blue points correspond to (anti)$^{3}$He with TOF information; other (anti)particles are shown in black. \textbf{(c)} Experimental results for the raw ratio of \ahe\ to \he\ in pp collisions at $\sqrt{s} = 13$ TeV as a function of rigidity. The vertical lines and boxes represent statistical and systematic uncertainties in terms of standard deviations, respectively. The black and red lines show the results from the Monte Carlo simulations with varied \sigmainel. \textbf{(d)} Experimental ratio of \ahe\ with TOF information over \ahe\ reconstructed in the TPC in the 10\% most central Pb--Pb collisions at $\sqrt{s_{\rm NN}} = 5.02$ TeV as a function of rigidity. The black and red lines show the results from the Monte Carlo simulations with varied \sigmainel. \textbf{(e)} The raw ratio of \ahe\ to \he\ in a particular rigidity interval as a function of \sigmainel\ for $\langle A\rangle = 17.4$. The fit to the results from Monte Carlo simulations (black points) shows the dependence of the observable on \sigmainel\ according to the Lambert--Beer formula. The horizontal dashed blue lines show the central value and $1\sigma$ uncertainties for the measured observable and their intersection with the Lambert--Beer function determines \sigmainel\ limits (yellow lines). \textbf{(f)} Extraction of \sigmainel\ for $\langle A\rangle = 34.7$ analogous to the panel \textbf{e}, with \sigmainel\ limits shown as magenta lines.}
\label{fig:Scheme1}
\end{figure}

The measurement of the inelastic cross sections under controlled conditions requires a beam with a well-defined momentum and a target whose material and its spatial distribution are well-known. Since no \ahe\ beams are available, we exploit the antimatter production at the LHC and the excellent identification and momentum determination for \ahe\ in ALICE as an equivalent setup. In our study, the ALICE detector itself serves as a target for the inelastic processes. A detailed description of the detector and its performance is available in Refs.~\cite{Aamodt:2008zz,Abelev:2014ffa}. Here, serving as probes, \ahe\ and \he\ nuclei are produced in pp and Pb--Pb collisions. At LHC high energies, \ahe\ and \he\ are produced in same amounts on average. The primordial ratio can be derived from precise antiproton-to-proton measurements~\cite{Aamodt:2010dx,Abbas:2013rua} and in pp collisions at the centre-of-mass energy of $\sqrt{s} = $ 13 TeV corresponds to $0.994 \pm 0.045$. 
The ALICE subdetectors that are considered as targets are the Inner Tracking System (ITS), the Time Projection Chamber (TPC) and the Transition Radiation Detector (TRD). A schematic representation of the ALICE detector is shown in Fig.~\ref{fig:Scheme1}a. The material composition of the three subdetectors is diverse.
The detailed knowledge of the detector geometry and composition~\cite{TRDNote,Abelev:2014ffa} (see the Supplemental Material of Ref.~\cite{Acharya:2020cee} for a cumulative distribution of of the material in the ALICE apparatus) enables
the determination of the effective target material for this layered configuration (\nameref{sec:Methods}). Here \sigmainel\ can be estimated for three effective targets.
The first one is characterised by the average material of the ITS+TPC systems (with averaged atomic mass and charge numbers of $\langle A \rangle =$ 17.4 and $\langle Z \rangle =$ 8.5), the second one corresponds to the ITS+TPC+TRD systems ($\langle A \rangle =$ 31.8 and $\langle Z \rangle =$ 14.8)~\cite{Acharya:2020cee}, and the third one corresponds to the TRD system only ($\langle A \rangle =$ 34.7 and $\langle Z \rangle =$ 16.1). The values are obtained by weighting the contribution from different materials with their density times length crossed by particles. 

Figure~\ref{fig:Scheme1} shows a schematic representation of the
analysis steps necessary to extract \sigmainel. 
Figure~\ref{fig:Scheme1}a shows \ahe\ and \he\ tracks crossing the ALICE detector, with the annihilation occurring for the \ahe.
The momentum $p$ is measured via the determination of the track trajectory and curvature radius in the ALICE magnetic field ($B = 0.5$ T). Here \ahe\ and \he\ are first identified when they reach the TPC by the measurement of their specific energy loss (d$E$/d$x$) in the detector gas. The excellent separation power of this measurement is shown in Fig.~\ref{fig:Scheme1}b, where the d$E$/d$x$ is presented as a function of the particle rigidity ($p/z$) and $z$ denotes the charge of the particle crossing the TPC in units of the electron charge. 
Here the red dots represent all the nuclei that are reconstructed in the TPC, whereas the blue dots show the nuclei that survive up to the time-of-flight (TOF) detector where they are matched to a TOF hit.
A more detailed description of the employed particle identification methods can be found in \nameref{sec:Methods}.

We use two methods to evaluate \sigmainel. The first method, applied to pp data sample at $\sqrt{s} =\, 13$ TeV, relies on the comparison of the measured \ahe\ and \he\ yields (antibaryon-to-baryon method). In this case, the experimental observable is constituted by the reconstructed \ahe/\he\ ratio analogously to the method used elsewhere~\cite{Acharya:2020cee} for (anti)deuterons. The inelastic process that takes place in the ITS, TPC or TRD material manifests itself by the fact that fewer \ahe\ than \he\ candidates are detected (Fig.~\ref{fig:Scheme1}c). Both destructive and non-destructive inelastic processes contribute to this effect. Here the full circular blue symbols show the momentum-dependent \ahe/\he\ ratio measured in pp collisions as a function of the particle rigidity reconstructed at the primary vertex ($p_{\rm primary} / |z|$). The discontinuity of the \ahe/\he\ ratio observed at $p_{\rm primary} / |z| = 1$ GeV/$c$ is due to the additional requirement of a hit in the TOF detector for momenta above this value. This ratio can also be evaluated by means of a full-scale Monte Carlo (MC) simulation of antinuclei and nuclei traversing the ALICE detector.

The measured observables are compared in each momentum interval with simulations where \sigmainel\ is varied to obtain the inelastic cross sections.
We performed several full-scale simulations with variations in \sigmainel\ with respect to the standard parameterization implemented in the \geant package~\cite{Agostinelli:2002hh,Uzhinsky:2011zz} (Fig.~\ref{fig:Scheme1}c).
Figure~\ref{fig:Scheme1}e presents the simulated ratio as a function of \sigmainel\ parametrized using the Lambert--Beer law~\cite{Lambert}. For each momentum interval, the uncertainties of \sigmainel\ are obtained by requiring an agreement at $\pm 1\sigma$ with the measured observables, where $\sigma$ represents the total experimental uncertainty (statistical and systematic uncertainties added in quadrature).

The second method, employed in the Pb--Pb data analysis at a centre-of-mass energy per nucleon pair $\sqrt{s_{\rm NN}} =\, 5.02$ TeV, measures the disappearance of \ahe\ nuclei in the TRD detector only (TOF-to-TPC method). The ratio of \ahe\ with TOF information to all \ahe\ candidates is considered as an experimental observable. Figure~\ref{fig:Scheme1}d shows the momentum-dependent ratio of \ahe\ with a reconstructed TOF hit to all the \ahe\ candidates extracted from Pb--Pb collisions. As with the first method, this observable is also evaluated by means of a full-scale MC \geant simulation assuming different \sigmainel\ values.
Figure~\ref{fig:Scheme1}f shows the extraction of \sigmainel\ and its related uncertainties for one rigidity interval following the same procedure as the one used in the first method.

\begin{figure}[htbp]
\centering
\includegraphics[width=0.48\textwidth]{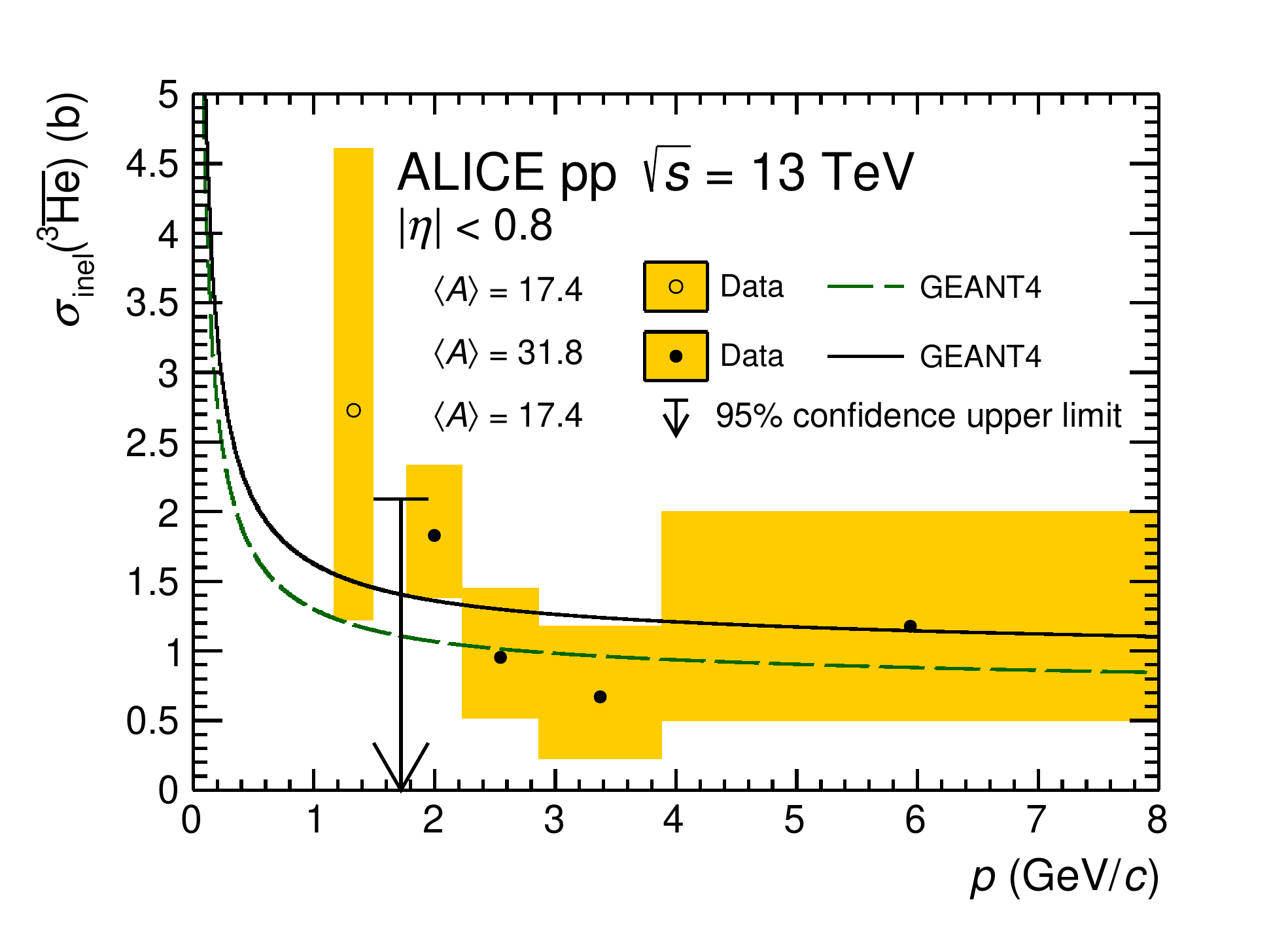}
\includegraphics[width=0.48\textwidth]{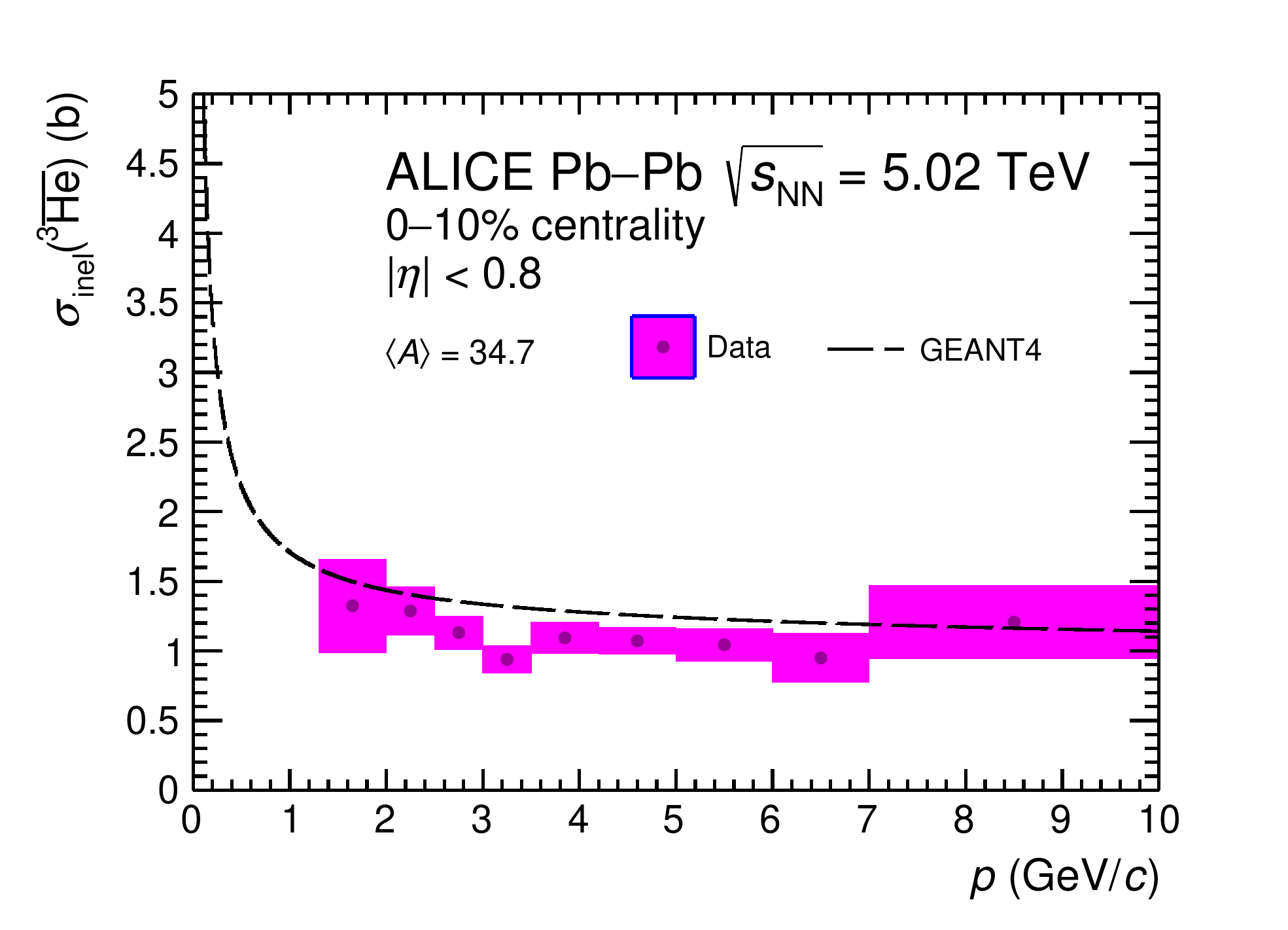}
\caption{Results for \sigmainel\ as a function of \ahe\ momentum. Results obtained from pp collisions at $\sqrt{s} =$ 13 TeV (left); results from the 10\% most central Pb--Pb collisions at $\sqrt{s_{\mathrm{NN}}} =\, 5.02$ TeV (right). The curves represent the \geant cross sections corresponding to the effective material probed by the different analyses. The arrow on the left plot shows the 95\% confidence limit on \sigmainel\ for $\langle A \rangle =$ 17.4. The different values of $\langle A \rangle$ correspond to the three different effective targets (see the main text for details). All the indicated uncertainties represent standard deviations.}
\label{fig:CrossSecwithG4param}
\end{figure}

The final results are shown in Fig.~\ref{fig:CrossSecwithG4param}. Figure~\ref{fig:CrossSecwithG4param} (left) shows the \sigmainel\ results from the pp data analysis with the yellow boxes representing the $\pm1\sigma$ uncertainty intervals. In  Fig.~\ref{fig:CrossSecwithG4param} (right), the histogram with the magenta error boxes shows \sigmainel\ extracted from the Pb--Pb data analysis. The results are shown as a function of the momentum $p$ at which the inelastic interaction occurs. Due to continuous energy loss inside the detector material, this momentum is lower than $p_{\rm primary}$ reconstructed at the primary vertex (\nameref{sec:Methods}). 
The antibaryon-to-baryon ratio method is applied in the pp data analysis, enabling the measurement of \sigmainel\ down to a low momentum. The copious background makes this method inapplicable in Pb--Pb collisions below $p = 1.5$ GeV/$c$ (\nameref{sec:Methods}). The TOF-to-TPC method is unavailable in this momentum range since \ahe\ nuclei don't reach the TOF due to the large energy loss and bending within the magnetic field.
On the other hand, for momentum values larger than $p = 1.5$ GeV/$c$, the yield of produced \ahe\ is substantially larger in Pb--Pb collisions, thus leading to higher statistical precision for this colliding system using the TOF-to-TPC method. The evaluation of systematic uncertainties is described in \nameref{sec:Methods}.
These two independent analysis methods, therefore, provide access to slightly different momentum ranges and to different $\langle A \rangle$ values and deliver consistent results in the common momentum region.

The cross section used by \geant for the average mass number $\langle A \rangle$ of the material is shown by the dashed lines in Fig.~\ref{fig:CrossSecwithG4param}. It is obtained from a Glauber model parameterization~\cite{Uzhinsky:2011zz} of the collisions of \ahe\ with target nuclei in which the antinucleon--nucleon cross section value is taken from measured $\overline{\rm p}$p collisions~\cite{Cudell:2001pn}. Agreement with the experimental \sigmainel\ is observed within two standard deviations in the studied momentum range.

\section{Propagation of antinuclei in the interstellar medium}

To estimate the transparency of our Galaxy to \ahe\ nuclei, we consider two examples of \ahe\ production sources. Results from another work~\cite{Shukla:2020bql} are used as input for the production cross section of \ahe\ from cosmic-ray collisions with interstellar medium.
As a DM source of \ahe\ we consider weakly interacting massive particle candidates with a mass of $100$ GeV/$c^{2}$ annihilating into $W^+W^-$ pairs followed by hadronization into (anti)nuclei~\cite{Carlson:2014ssa}.
In both cases, the yields of produced \ahe\ are determined employing the coalescence model that builds antinuclei from antineutrons and antiprotons that are close-by in phase space~\cite{Butler:1963pp,Scheibl:1998tk,ALICE:2015wav}.
More details about the cosmic-ray and DM sources are discussed in \nameref{sec:Methods}. Additional \ahe\ sources such as supernovae remnants~\cite{Tomassetti:2017izg}, antistars~\cite{Persic:1995ru,Poulin:2018wzu} and primordial black holes~\cite{Herms:2016vop,Barrau:2002mc,Serksnyte:2022onw} have not been included in this work.

\begin{figure}[htbp]
\centering
\includegraphics[width=0.9\textwidth]{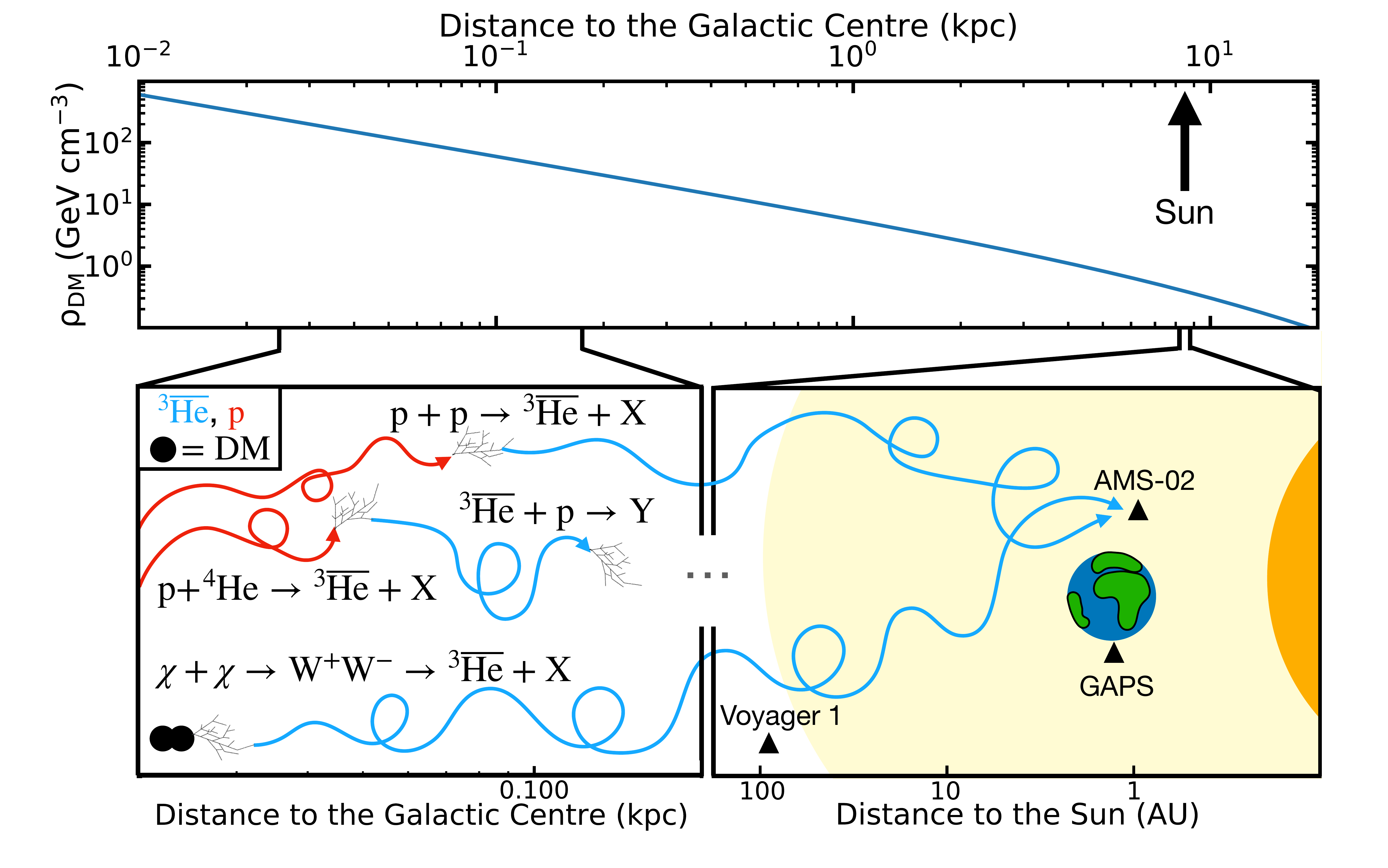}
\caption{Schematic of \ahe\ production and propagation in our Galaxy. Distribution of DM density $\rho_{\rm DM}$ in our Galaxy as a function of the distance from the galactic centre according to the Navarro–Frenk–White profile~\cite{Navarro:1995iw} (top). Graphical illustration of the \ahe\ production from cosmic-ray interactions with interstellar gas or DM ($\chi$) annihilations (bottom). The yellow halo represents the heliosphere and the Earth, Sun and positions of the Voyager 1, AMS-02 and GAPS experiments are depicted, too.}
\label{fig:PropagationChain}
\end{figure}

We consider the DM density distribution in our Galaxy according to the Navarro–Frenk–White profile~\cite{Navarro:1995iw} (Fig.~\ref{fig:PropagationChain} top), where a schematic of the \ahe\ production from cosmic-ray interaction with the interstellar gas or DM annihilations is also shown. 

The propagation of charged particles within galaxies is driven by magnetic fields.
The propagation is commonly described by a transport equation that includes the following terms:
(1) a source function, (2) diffusion, (3) convection, (4) momentum variations due to Coulomb scattering, diffusion and ionization processes, (5) fragmentation, decays and inelastic interactions.
This equation, discussed in more details in \nameref{sec:Methods}, can be solved numerically employing several propagation models~\cite{Kissmann:2017,Kissmann:2014sia,Evoli:2008,Strong:1998pw}. In this work the publicly available GALPROP code~\cite{Strong:1998pw} is employed. In the context of this calculation, our Galaxy is approximated by a cylindrical disk filled with an interstellar gas composed of hydrogen ($ \approx 90\%$) and $^4$He ($ \approx 10\%$) with an average hydrogen number density of $\sim 1$ atom/cm$^3$~\cite{Moskalenko:2001ya}.
The gas distribution within our Galaxy is constrained by several astronomical spectroscopy measurements~\cite{Bronfman:1988,Gordon:1976,Cordes:1991,Dickey:1990}.
GALPROP provides the propagation of particles up to the boundaries of the Solar System. To estimate the particle flux inside the Solar System, the effect of the solar magnetic field must be taken into account. This can be achieved by employing the Force Field approximation or dedicated models like HelMod~\cite{Gleeson:1968zza,Boschini:2019ubh}.
The whole propagation chain is benchmarked using several species of cosmic rays, including protons and light nuclei (up to $Z =\, 28$)~\cite{Boschini:2020jty}. The cosmic-ray injection spectra and the propagation parameters are tuned to match the measurements of protons and light nuclei both outside~\cite{Cummings:2016pdr} and within the Solar System~\cite{Aguilar:2015ooa,Engelmann:1990zz,Ahn:2008my}.

After their production, the \ahe\ nuclei need to travel a distance of several kiloparcecs to reach Earth~\cite{Boschini:2020jty,Navarro:1995iw}. During this passage, they might encounter protons or $^4\mathrm{He}$ nuclei in the interstellar gas and inelastically interact. Non-destructive inelastic processes can occur and cause a substantial energy loss that results in a so-called tertiary \ahe\ source peaked at low kinetic energies. Such a tertiary source component however only contributes a few percent of the total flux~\cite{Korsmeier:2017xzj, vonDoetinchem:2020vbj}. We neglect this small contribution because we cannot distinguish between destructive and non-destructive inelastic processes. To model the total cross section of inelastic processes, we scale the momentum-dependent \geant parameterization of the \ahe--p inelastic cross section with the correction factors obtained from our measurements. 
For the low-momentum range ($1.17\leq p<1.5$ GeV/$c$) we consider the results from pp collisions and for the high-momentum range ($1.5\leq p<10$ GeV/$c$), results from Pb--Pb collisions. The correction factors from the ALICE measurements and their uncertainties are parametrized with a continuous function employing a combination of polynomial and exponential functions. The additional uncertainty due to scaling with $A$ is estimated to be lower than 8\%~\cite{Uzhinsky:2011zz} (\nameref{sec:Methods}). For the extrapolation to momenta above the measured momentum range, we consider the correction factor corresponding to the last measured momentum interval (Fig.~\ref{fig:CrossSecwithG4param} right).
The resulting \ahe-p inelastic cross section as a function of the \ahe\ kinetic energy per nucleon is shown in \nameref{sec:ExtendedData} Fig.~\ref{fig:CrossSectionScaled} together with the \geant parameterization and the model employed in another work~\cite{Korsmeier:2017xzj}. The same procedure is applied to describe the \ahe--\heFour\ inelastic processes.
These scaled inelastic cross sections have been implemented in GALPROP.

\begin{figure}[htbp]
\centering
\includegraphics[width=0.49\textwidth]{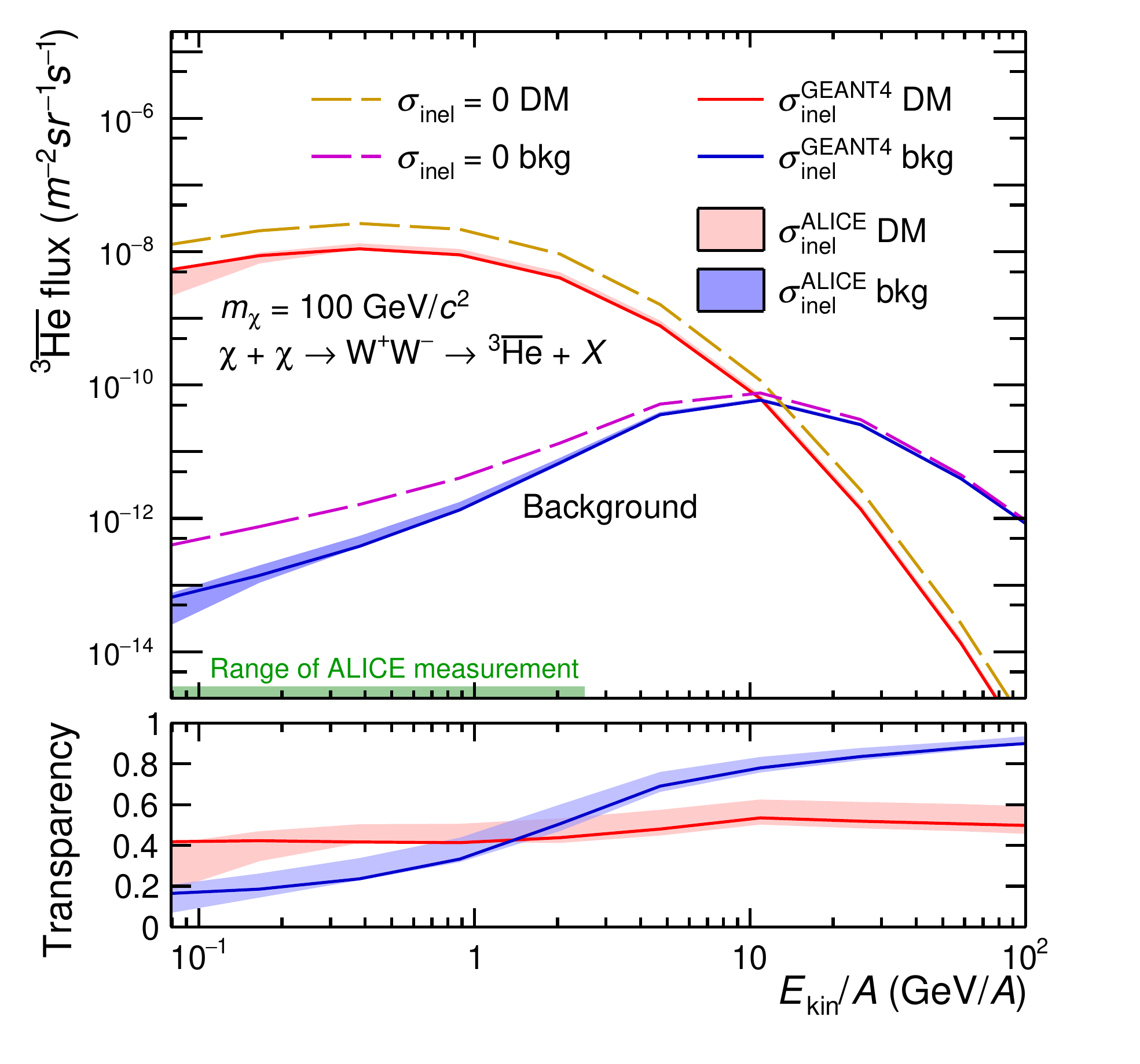}
\includegraphics[width=0.49\textwidth]{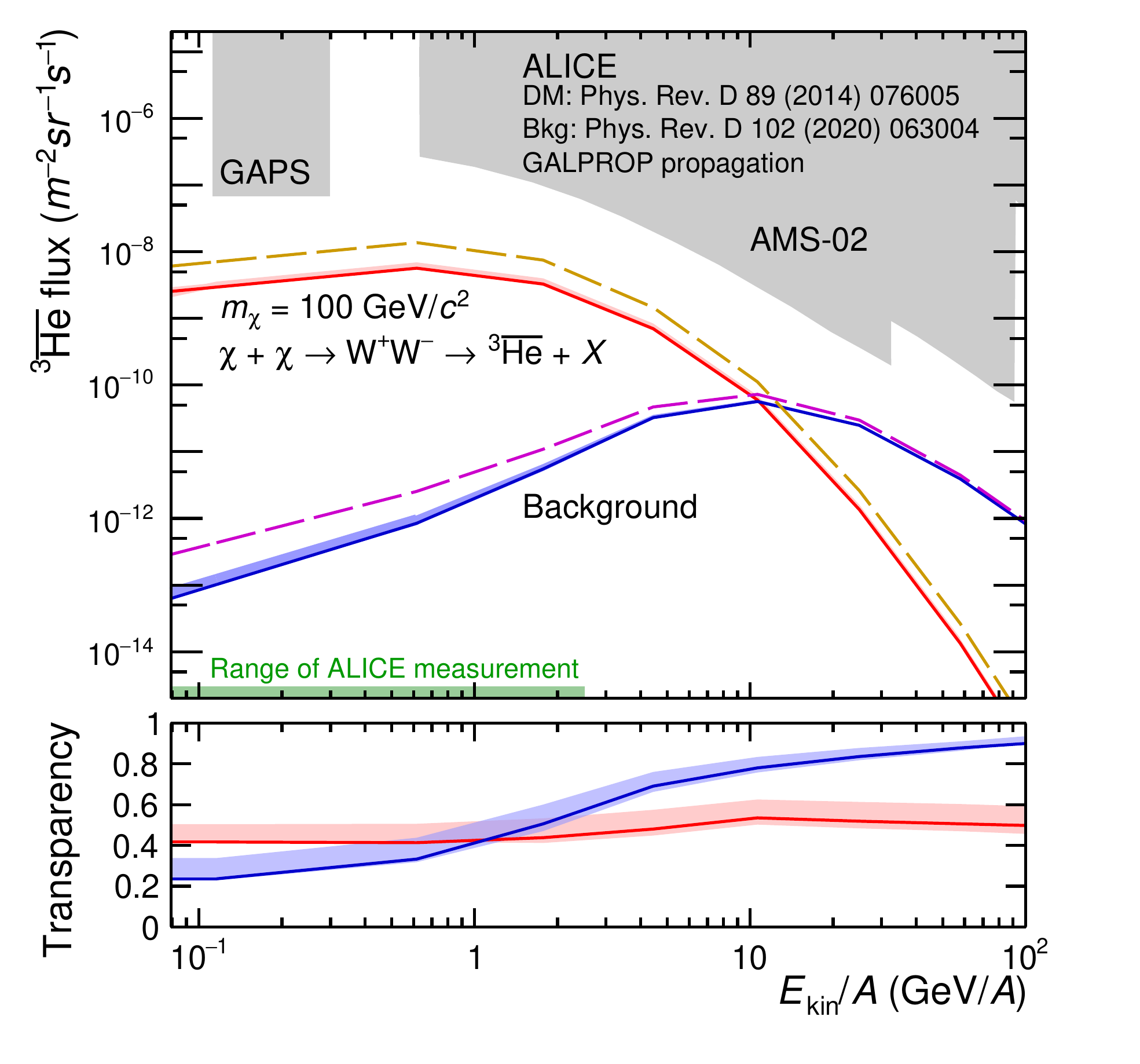}
\caption{Expected \ahe\ flux near Earth before and after solar modulation. Data before (left) and after (right) solar modulation. The latter is obtained using the force-field method with modulation potential $\mathrm{\phi} =$ 400 MV. The results are shown as a function of kinetic energy per nucleon $E_{\rm kin} / A$. Fluxes for DM signal $\chi$ (red) and cosmic-ray background (blue) antihelium nuclei for different cases of inelastic cross section used in the calculations (top). The bands show the results obtained with \sigmainel\ from ALICE measurements, and the full lines correspond to the results using the \geant\ parameterizations. The dashed lines show the fluxes obtained with \sigmainel\ set to zero for DM signal (orange line) and for cosmic-ray background (magenta line). The green band on the x axis indicates the kinetic energy range corresponding to the ALICE measurement for \sigmainel. Transparency of our Galaxy to the propagation of \ahe\ outside (left) and inside (right) the Solar System (bottom). The shaded areas (top right) show the expected sensitivity of the GAPS~\cite{GAPS:2020axg} and AMS-02~\cite{Korsmeier:2017xzj} experiments. The top panels also shows the fluxes obtained with \sigmainel\ set to zero. Only the uncertainties relative to the measured \sigmainel\ are shown, which represent standard deviations. The calculations employ the \ahe\ DM source described elsewhere~\cite{Carlson:2014ssa} and the \ahe\ production cross section from cosmic-ray background~\cite{Shukla:2020bql}.}
\label{results}
\end{figure}

The expected \ahe\ flux near Earth after all the propagation steps (\nameref{sec:Methods}) with and without the effect of solar modulations is shown in the right and left panels of Fig.~\ref{results}, respectively. Solar modulation is implemented using the force-field method~\cite{Gleeson:1968zza}.
The effect of inelastic interactions is demonstrated by showing the full propagation chain once with \sigmainel\ set to zero and once with the inelastic cross section extracted from the ALICE measurement. Only the uncertainties relative to the measured \sigmainel\ are propagated and presented in Fig.~\ref{results}.
The inelastic collisions of \ahe\ with the interstellar gas lead to a notable reduction in the expected flux for the signal candidates from DM as well as the background from cosmic-ray collisions.

The transparency of our Galaxy to the \ahe\ passage is defined by the ratio of the flux obtained with and without inelastic processes in GALPROP. The transparency values as a function of the kinetic energy obtained with \sigmainel\ from the \geant parameterization and from the ALICE measurements are shown in Fig.~\ref{results} (bottom) by the coloured lines and bands, respectively. The transparency profiles obtained with a solar modulation potential of 400 MV do not differ much from the non-modulated distributions (Fig.~\ref{results}, bottom left and right). This is because the solar modulation reshuffles the yield from the more abundant high-momentum range to 
lower energies, but all the transparency profiles are rather flat as a function of the particle energy. A transparency of the Galaxy of about 50\% is estimated for \ahe\ from the considered DM source~\cite{Carlson:2014ssa} and of about 25\% for low-energy \ahe\ from cosmic-ray interactions~\cite{Shukla:2020bql}. The latter increases further up to full transparency at higher energies. The different behaviour in the two cases is caused both by both different underlying spectral shape and different distribution of production points of the two sources, underlining the importance of full propagation studies (\nameref{sec:Methods}). The employment of an alternative set of propagation parameters from Ref.~\cite{Cuoco:2016eej} results in $40-60\%$ lower transparency at low $E_{\rm kin}$ than using the propagation parameters from Ref.~\cite{Boschini:2020jty} (\nameref{sec:Methods}).

The calculated \ahe\ transparency is found to be consistent -- within uncertainties -- with the \geant parameterization. It must be clearly noted that previously it was not possible to quantify the uncertainty of the parameterizations employed in \geant or proposed elsewhere~\cite{Korsmeier:2017xzj} due to the lack of experimental data.
To quantify the improvement originating from our study, we, therefore, simply compare the full difference between no inelastic interaction and the alternative parameterizations ($\sim 50$\% for the signal from DM and up to 75\% for background) to our newly established uncertainties of about 10\%--15\% after solar modulation. We have, thus, verified that the uncertainty related to nuclear absorption is subleading with respect to other possible contributions in the cosmic-ray and DM modelling, particularly the production mechanism and propagation description~\cite{Korsmeier:2017xzj,Carlson:2014ssa,Shukla:2020bql,vonDoetinchem:2020vbj}. Note that the propagation example provided in this work does not cover the full range of uncertainties related to \ahe\ flux modelling (\nameref{sec:Methods}), rather, it delivers a clear road map for future studies. The measured \sigmainel\ and the developed methodology can be employed to carry out the propagation of \ahe\ using any DM or cosmic-ray interaction modelling as a source. Since a large separation between signal and background is retained for low kinetic energies, our results clearly underline that the search for \ahe\ in space remains a very promising channel for the discovery of DM. These studies will be extended to $^4\overline{\mathrm{He}}$ and to the lower momentum region in the near future with much larger datasets that will be collected in the coming few years.

%TC:ignore

\newenvironment{acknowledgement}{\relax}{\relax}
\begin{acknowledgement}
\section*{Acknowledgements}

We thank A. Strong for his guidance in the implementation of the antinuclei propagation within the GALPROP code, A. Shukla and P. von Doetinchem for providing the $^{3}\overline{\rm He}$ production cross sections in cosmic-ray collisions with interstellar medium, and J. Herms and A. Ibarra for model calculations of the $^{3}\overline{\rm He}$ spectra stemming from DM annihilation.

% Version: 2021-12-14

The ALICE Collaboration would like to thank all its engineers and technicians for their invaluable contributions to the construction of the experiment and the CERN accelerator teams for the outstanding performance of the LHC complex.
The ALICE Collaboration gratefully acknowledges the resources and support provided by all Grid centres and the Worldwide LHC Computing Grid (WLCG) collaboration.
The ALICE Collaboration acknowledges the following funding agencies for their support in building and running the ALICE detector:
A. I. Alikhanyan National Science Laboratory (Yerevan Physics Institute) Foundation (ANSL), State Committee of Science and World Federation of Scientists (WFS), Armenia;
Austrian Academy of Sciences, Austrian Science Fund (FWF): [M 2467-N36] and Nationalstiftung f\"{u}r Forschung, Technologie und Entwicklung, Austria;
Ministry of Communications and High Technologies, National Nuclear Research Center, Azerbaijan;
Conselho Nacional de Desenvolvimento Cient\'{\i}fico e Tecnol\'{o}gico (CNPq), Financiadora de Estudos e Projetos (Finep), Funda\c{c}\~{a}o de Amparo \`{a} Pesquisa do Estado de S\~{a}o Paulo (FAPESP) and Universidade Federal do Rio Grande do Sul (UFRGS), Brazil;
Ministry of Education of China (MOEC) , Ministry of Science \& Technology of China (MSTC) and National Natural Science Foundation of China (NSFC), China;
Ministry of Science and Education and Croatian Science Foundation, Croatia;
Centro de Aplicaciones Tecnol\'{o}gicas y Desarrollo Nuclear (CEADEN), Cubaenerg\'{\i}a, Cuba;
Ministry of Education, Youth and Sports of the Czech Republic, Czech Republic;
The Danish Council for Independent Research | Natural Sciences, the VILLUM FONDEN and Danish National Research Foundation (DNRF), Denmark;
Helsinki Institute of Physics (HIP), Finland;
Commissariat \`{a} l'Energie Atomique (CEA) and Institut National de Physique Nucl\'{e}aire et de Physique des Particules (IN2P3) and Centre National de la Recherche Scientifique (CNRS), France;
Bundesministerium f\"{u}r Bildung und Forschung (BMBF) and GSI Helmholtzzentrum f\"{u}r Schwerionenforschung GmbH, Germany;
General Secretariat for Research and Technology, Ministry of Education, Research and Religions, Greece;
National Research, Development and Innovation Office, Hungary;
Department of Atomic Energy Government of India (DAE), Department of Science and Technology, Government of India (DST), University Grants Commission, Government of India (UGC) and Council of Scientific and Industrial Research (CSIR), India;
Indonesian Institute of Science, Indonesia;
Istituto Nazionale di Fisica Nucleare (INFN), Italy;
Japanese Ministry of Education, Culture, Sports, Science and Technology (MEXT) and Japan Society for the Promotion of Science (JSPS) KAKENHI, Japan;
Consejo Nacional de Ciencia (CONACYT) y Tecnolog\'{i}a, through Fondo de Cooperaci\'{o}n Internacional en Ciencia y Tecnolog\'{i}a (FONCICYT) and Direcci\'{o}n General de Asuntos del Personal Academico (DGAPA), Mexico;
Nederlandse Organisatie voor Wetenschappelijk Onderzoek (NWO), Netherlands;
The Research Council of Norway, Norway;
Commission on Science and Technology for Sustainable Development in the South (COMSATS), Pakistan;
Pontificia Universidad Cat\'{o}lica del Per\'{u}, Peru;
Ministry of Education and Science, National Science Centre and WUT ID-UB, Poland;
Korea Institute of Science and Technology Information and National Research Foundation of Korea (NRF), Republic of Korea;
Ministry of Education and Scientific Research, Institute of Atomic Physics, Ministry of Research and Innovation and Institute of Atomic Physics and University Politehnica of Bucharest, Romania;
Joint Institute for Nuclear Research (JINR), Ministry of Education and Science of the Russian Federation, National Research Centre Kurchatov Institute, Russian Science Foundation and Russian Foundation for Basic Research, Russia;
Ministry of Education, Science, Research and Sport of the Slovak Republic, Slovakia;
National Research Foundation of South Africa, South Africa;
Swedish Research Council (VR) and Knut \& Alice Wallenberg Foundation (KAW), Sweden;
European Organization for Nuclear Research, Switzerland;
Suranaree University of Technology (SUT), National Science and Technology Development Agency (NSDTA), Suranaree University of Technology (SUT), Thailand Science Research and Innovation (TSRI) and National Science, Research and Innovation Fund (NSRF), Thailand;
Turkish Energy, Nuclear and Mineral Research Agency (TENMAK), Turkey;
National Academy of  Sciences of Ukraine, Ukraine;
Science and Technology Facilities Council (STFC), United Kingdom;
National Science Foundation of the United States of America (NSF) and United States Department of Energy, Office of Nuclear Physics (DOE NP), United States of America.    %%%%%%% done by webmaster team

In addition, individual groups or members have received support from the German Research Foundation (DFG), Germany

\end{acknowledgement}

%%%%%%%% Bibliography (In case of using bibtex generate the bbl requested by arXiv)
%\bibliographystyle{alpha}   % Put here the style file name for the paper, i.e.apsrev4-1, utphys
%\bibliography{biblio}
\bibliographystyle{utphys} 
\bibliography{main}

\providecommand{\href}[2]{#2}\begingroup\raggedright\begin{thebibliography}{10}

\bibitem{SimonGillo:1995dh}
{\bfseries NA44} Collaboration, J.~Simon-Gillo {\em et~al.}, ``{Deuteron and
  anti-deuteron production in CERN experiment NA44}'',
  \href{http://dx.doi.org/10.1016/0375-9474(95)00259-4}{{\em Nucl. Phys. A}
  {\bfseries 590} (1995) 483C--486C}.

\bibitem{Armstrong:2000gd}
{\bfseries E864} Collaboration, T.~Armstrong {\em et~al.}, ``{Anti-deuteron
  yield at the AGS and coalescence implications}'',
  \href{http://dx.doi.org/10.1103/PhysRevLett.85.2685}{{\em Phys. Rev. Lett.}
  {\bfseries 85} (2000) 2685--2688},
  \href{http://arxiv.org/abs/nucl-ex/0005001}{{\ttfamily
  arXiv:nucl-ex/0005001}}.

\bibitem{Afanasev:2000ku}
{\bfseries NA49} Collaboration, S.~Afanasiev {\em et~al.}, ``{Deuteron
  production in central Pb + Pb collisions at 158-A-GeV}'',
  \href{http://dx.doi.org/10.1016/S0370-2693(00)00746-2}{{\em Phys. Lett. B}
  {\bfseries 486} (2000) 22--28}.

\bibitem{Anticic:2004yj}
{\bfseries NA49} Collaboration, T.~Anticic {\em et~al.}, ``{Energy and
  centrality dependence of deuteron and proton production in Pb + Pb collisions
  at relativistic energies}'',
  \href{http://dx.doi.org/10.1103/PhysRevC.69.024902}{{\em Phys. Rev. C}
  {\bfseries 69} (2004) 024902}.

\bibitem{Adler:2004uy}
{\bfseries PHENIX} Collaboration, S.~Adler {\em et~al.}, ``{Deuteron and
  antideuteron production in Au + Au collisions at $\sqrt{s_{\rm NN}}$ = 200
  GeV}'', \href{http://dx.doi.org/10.1103/PhysRevLett.94.122302}{{\em Phys.
  Rev. Lett.} {\bfseries 94} (2005) 122302},
  \href{http://arxiv.org/abs/nucl-ex/0406004}{{\ttfamily
  arXiv:nucl-ex/0406004}}.

\bibitem{Alper:1973my}
B.~Alper {\em et~al.}, ``{Large angle production of stable particles heavier
  than the proton and a search for quarks at the cern intersecting storage
  rings}'', \href{http://dx.doi.org/10.1016/0370-2693(73)90700-4}{{\em Phys.
  Lett. B} {\bfseries 46} (1973) 265--268}.

\bibitem{Henning:1977mt}
{\bfseries British-Scandinavian-MIT} Collaboration, S.~Henning {\em et~al.},
  ``{Production of Deuterons and anti-Deuterons in Proton Proton Collisions at
  the CERN ISR}'', \href{http://dx.doi.org/10.1007/BF02822248}{{\em Lett. Nuovo
  Cim.} {\bfseries 21} (1978) 189}.

\bibitem{Alexopoulos:2000jk}
T.~Alexopoulos {\em et~al.}, ``{Cross-sections for deuterium, tritium, and
  helium production in $\bar{p}p$ collisions at $\sqrt{s}$ = 1.8 TeV}'',
  \href{http://dx.doi.org/10.1103/PhysRevD.62.072004}{{\em Phys. Rev. D}
  {\bfseries 62} (2000) 072004}.

\bibitem{Aktas:2004pq}
{\bfseries H1} Collaboration, A.~Aktas {\em et~al.}, ``{Measurement of
  anti-deuteron photoproduction and a search for heavy stable charged particles
  at HERA}'', \href{http://dx.doi.org/10.1140/epjc/s2004-01978-x}{{\em Eur.
  Phys. J. C} {\bfseries 36} (2004) 413--423},
  \href{http://arxiv.org/abs/hep-ex/0403056}{{\ttfamily arXiv:hep-ex/0403056}}.

\bibitem{Asner:2006pw}
{\bfseries CLEO} Collaboration, D.~Asner {\em et~al.}, ``{Anti-deuteron
  production in $\Upsilon(nS)$ decays and the nearby continuum}'',
  \href{http://dx.doi.org/10.1103/PhysRevD.75.012009}{{\em Phys. Rev. D}
  {\bfseries 75} (2007) 012009},
  \href{http://arxiv.org/abs/hep-ex/0612019}{{\ttfamily arXiv:hep-ex/0612019}}.

\bibitem{Schael:2006fd}
{\bfseries ALEPH} Collaboration, S.~Schael {\em et~al.}, ``{Deuteron and
  anti-deuteron production in e$^{+}$e$^{-}$ collisions at the Z resonance}'',
  \href{http://dx.doi.org/10.1016/j.physletb.2006.06.043}{{\em Phys. Lett. B}
  {\bfseries 639} (2006) 192--201},
  \href{http://arxiv.org/abs/hep-ex/0604023}{{\ttfamily arXiv:hep-ex/0604023}}.

\bibitem{ALICE:2015wav}
{\bfseries ALICE} Collaboration, J.~Adam {\em et~al.}, ``{Production of light
  nuclei and anti-nuclei in pp and Pb-Pb collisions at energies available at
  the CERN Large Hadron Collider}'',
  \href{http://dx.doi.org/10.1103/PhysRevC.93.024917}{{\em Phys. Rev. C}
  {\bfseries 93} (2016) 024917},
  \href{http://arxiv.org/abs/1506.08951}{{\ttfamily arXiv:1506.08951
  [nucl-ex]}}.

\bibitem{Adam:2015yta}
{\bfseries ALICE} Collaboration, J.~Adam {\em et~al.},
  ``{$^{3}_{\Lambda}\mathrm H$ and $^{3}_{\bar{\Lambda}} \overline{\mathrm H}$
  production in Pb--Pb collisions at $\sqrt{s_{\rm NN}} =$ 2.76 TeV}'',
  \href{http://dx.doi.org/10.1016/j.physletb.2016.01.040}{{\em Phys. Lett.}
  {\bfseries B754} (2016) 360--372},
\href{http://arxiv.org/abs/1506.08453}{{\ttfamily arXiv:1506.08453 [nucl-ex]}}.
%%CITATION = ARXIV:1506.08453;%%.

\bibitem{Acharya:2017dmc}
{\bfseries ALICE} Collaboration, S.~Acharya {\em et~al.}, ``{Measurement of
  deuteron spectra and elliptic flow in Pb--Pb collisions at $\sqrt{s_{\mathrm
  {NN}}}$ = 2.76 TeV at the LHC}'',
  \href{http://dx.doi.org/10.1140/epjc/s10052-017-5222-x}{{\em Eur. Phys. J.}
  {\bfseries C77} (2017) 658},
\href{http://arxiv.org/abs/1707.07304}{{\ttfamily arXiv:1707.07304 [nucl-ex]}}.
%%CITATION = ARXIV:1707.07304;%%.

\bibitem{Acharya:2019rgc}
{\bfseries ALICE} Collaboration, S.~Acharya {\em et~al.}, ``{Multiplicity
  dependence of (anti-)deuteron production in pp collisions at $\sqrt{s}$ = 7
  TeV}'', \href{http://dx.doi.org/10.1016/j.physletb.2019.05.028}{{\em Phys.
  Lett.} {\bfseries B794} (2019) 50--63},
\href{http://arxiv.org/abs/1902.09290}{{\ttfamily arXiv:1902.09290 [nucl-ex]}}.
%%CITATION = ARXIV:1902.09290;%%.

\bibitem{Agakishiev:2011ib}
{\bfseries STAR} Collaboration, H.~Agakishiev {\em et~al.}, ``{Observation of
  the antimatter helium-4 nucleus}'',
  \href{http://dx.doi.org/10.1038/nature10079}{{\em Nature} {\bfseries 473}
  (2011) 353},
\href{http://arxiv.org/abs/1103.3312}{{\ttfamily arXiv:1103.3312 [nucl-ex]}}.
%%CITATION = ARXIV:1103.3312;%%.

\bibitem{Abelev:2010}
{\bfseries STAR} Collaboration, B.~I. Abelev {\em et~al.}, ``{Observation of an
  Antimatter Hypernucleus}'',
  \href{http://dx.doi.org/10.1126/science.1183980}{{\em Science} {\bfseries
  328} (2010) 58--62},
\href{http://arxiv.org/abs/1003.2030}{{\ttfamily arXiv:1003.2030 [nucl-ex]}}.
%%CITATION = 1003.2030;%%.

\bibitem{STAR:2001pbk}
{\bfseries STAR} Collaboration, C.~Adler {\em et~al.}, ``{Anti-deuteron and
  anti-He-3 production in s(NN)**(1/2) = 130-GeV Au+Au collisions}'',
  \href{http://dx.doi.org/10.1103/PhysRevLett.87.262301}{{\em Phys. Rev. Lett.}
  {\bfseries 87} (2001) 262301},
  \href{http://arxiv.org/abs/nucl-ex/0108022}{{\ttfamily
  arXiv:nucl-ex/0108022}}. [Erratum: Phys.Rev.Lett. 87, 279902 (2001)].

\bibitem{Adam:2015pna}
{\bfseries ALICE} Collaboration, J.~Adam {\em et~al.}, ``{Precision measurement
  of the mass difference between light nuclei and anti-nuclei}'',
  \href{http://dx.doi.org/10.1038/nphys3432}{{\em Nature Phys.} {\bfseries 11}
  (2015) 811--814}, \href{http://arxiv.org/abs/1508.03986}{{\ttfamily
  arXiv:1508.03986 [nucl-ex]}}.

\bibitem{Adam:2019phl}
{\bfseries STAR} Collaboration, J.~Adam {\em et~al.}, ``{Measurement of the
  mass difference and the binding energy of the hypertriton and
  antihypertriton}'', \href{http://dx.doi.org/10.1038/s41567-020-0799-7}{{\em
  Nature Phys.} {\bfseries 16} (2020) 409--412},
  \href{http://arxiv.org/abs/1904.10520}{{\ttfamily arXiv:1904.10520
  [hep-ex]}}.

\bibitem{STAR:2015kha}
{\bfseries STAR} Collaboration, L.~Adamczyk {\em et~al.}, ``{Measurement of
  Interaction between Antiprotons}'',
  \href{http://dx.doi.org/10.1038/nature15724}{{\em Nature} {\bfseries 527}
  (2015) 345--348}, \href{http://arxiv.org/abs/1507.07158}{{\ttfamily
  arXiv:1507.07158 [nucl-ex]}}.

\bibitem{Ibarra:2012cc}
A.~Ibarra and S.~Wild, ``{Prospects of antideuteron detection from dark matter
  annihilations or decays at AMS-02 and GAPS}'',
  \href{http://dx.doi.org/10.1088/1475-7516/2013/02/021}{{\em JCAP} {\bfseries
  02} (2013) 021}, \href{http://arxiv.org/abs/1209.5539}{{\ttfamily
  arXiv:1209.5539 [hep-ph]}}.

\bibitem{Persic:1995ru}
M.~Persic, P.~Salucci, and F.~Stel, ``{The Universal rotation curve of spiral
  galaxies: 1. The Dark matter connection}'',
  \href{http://dx.doi.org/10.1093/mnras/278.1.27}{{\em Mon. Not. Roy. Astron.
  Soc.} {\bfseries 281} (1996) 27},
  \href{http://arxiv.org/abs/astro-ph/9506004}{{\ttfamily
  arXiv:astro-ph/9506004}}.

\bibitem{Poulin:2018wzu}
V.~Poulin, P.~Salati, I.~Cholis, M.~Kamionkowski, and J.~Silk, ``{Where do the
  AMS-02 antihelium events come from?}'',
  \href{http://dx.doi.org/10.1103/PhysRevD.99.023016}{{\em Phys. Rev. D}
  {\bfseries 99} (2019) 023016},
  \href{http://arxiv.org/abs/1808.08961}{{\ttfamily arXiv:1808.08961
  [astro-ph.HE]}}.

\bibitem{Aghanim:2018eyx}
{\bfseries Planck} Collaboration, N.~Aghanim {\em et~al.}, ``{Planck 2018
  results. VI. Cosmological parameters}'',
  \href{http://dx.doi.org/10.1051/0004-6361/201833910}{{\em Astron. Astrophys.}
  {\bfseries 641} (2020) A6}, \href{http://arxiv.org/abs/1807.06209}{{\ttfamily
  arXiv:1807.06209 [astro-ph.CO]}}.

\bibitem{Bond:1984fp}
J.~Bond and G.~Efstathiou, ``{Cosmic background radiation anisotropies in
  universes dominated by nonbaryonic dark matter}'',
  \href{http://dx.doi.org/10.1086/184362}{{\em Astrophys. J. Lett.} {\bfseries
  285} (1984) L45--L48}.

\bibitem{deBernardis:2000sbo}
{\bfseries Boomerang} Collaboration, P.~de~Bernardis {\em et~al.}, ``{A Flat
  universe from high resolution maps of the cosmic microwave background
  radiation}'', \href{http://dx.doi.org/10.1038/35010035}{{\em Nature}
  {\bfseries 404} (2000) 955--959},
  \href{http://arxiv.org/abs/astro-ph/0004404}{{\ttfamily
  arXiv:astro-ph/0004404}}.

\bibitem{Clowe:2006eq}
D.~Clowe, M.~Bradac, A.~H. Gonzalez, M.~Markevitch, S.~W. Randall, C.~Jones,
  and D.~Zaritsky, ``{A direct empirical proof of the existence of dark
  matter}'', \href{http://dx.doi.org/10.1086/508162}{{\em Astrophys. J. Lett.}
  {\bfseries 648} (2006) L109--L113},
  \href{http://arxiv.org/abs/astro-ph/0608407}{{\ttfamily
  arXiv:astro-ph/0608407}}.

\bibitem{Carlson:2014ssa}
E.~Carlson, A.~Coogan, T.~Linden, S.~Profumo, A.~Ibarra, and S.~Wild,
  ``{Antihelium from Dark Matter}'',
  \href{http://dx.doi.org/10.1103/PhysRevD.89.076005}{{\em Phys. Rev. D}
  {\bfseries 89} (2014) 076005},
  \href{http://arxiv.org/abs/1401.2461}{{\ttfamily arXiv:1401.2461 [hep-ph]}}.

\bibitem{Korsmeier:2017xzj}
M.~Korsmeier, F.~Donato, and N.~Fornengo, ``{Prospects to verify a possible
  dark matter hint in cosmic antiprotons with antideuterons and antihelium}'',
  \href{http://dx.doi.org/10.1103/PhysRevD.97.103011}{{\em Phys. Rev.}
  {\bfseries D97} (2018) 103011},
\href{http://arxiv.org/abs/1711.08465}{{\ttfamily arXiv:1711.08465
  [astro-ph.HE]}}.
%%CITATION = ARXIV:1711.08465;%%.

\bibitem{vonDoetinchem:2020vbj}
P.~von Doetinchem {\em et~al.}, ``{Cosmic-ray antinuclei as messengers of new
  physics: status and outlook for the new decade}'',
  \href{http://dx.doi.org/10.1088/1475-7516/2020/08/035}{{\em JCAP} {\bfseries
  08} (2020) 035}, \href{http://arxiv.org/abs/2002.04163}{{\ttfamily
  arXiv:2002.04163 [astro-ph.HE]}}.

\bibitem{Winkler:2020ltd}
M.~W. Winkler and T.~Linden, ``{Dark Matter Annihilation Can Produce a
  Detectable Antihelium Flux through $\bar{\Lambda}_b$ Decays}'',
  \href{http://dx.doi.org/10.1103/PhysRevLett.126.101101}{{\em Phys. Rev.
  Lett.} {\bfseries 126} (2021) 101101},
  \href{http://arxiv.org/abs/2006.16251}{{\ttfamily arXiv:2006.16251
  [hep-ph]}}.

\bibitem{Bellini:2020cbj}
F.~Bellini, K.~Blum, A.~P. Kalweit, and M.~Puccio, ``{Examination of
  coalescence as the origin of nuclei in hadronic collisions}'',
  \href{http://dx.doi.org/10.1103/PhysRevC.103.014907}{{\em Phys. Rev. C}
  {\bfseries 103} (2021) 014907},
  \href{http://arxiv.org/abs/2007.01750}{{\ttfamily arXiv:2007.01750
  [nucl-th]}}.

\bibitem{Kachelriess:2020amp}
M.~Kachelriess, S.~Ostapchenko, and J.~Tjemsland, ``{On nuclear coalescence in
  small interacting systems}'',
  \href{http://dx.doi.org/10.1140/epja/s10050-021-00469-w}{{\em Eur. Phys. J.
  A} {\bfseries 57} (2021) 167},
  \href{http://arxiv.org/abs/2012.04352}{{\ttfamily arXiv:2012.04352
  [hep-ph]}}.

\bibitem{Braun-Munzinger:2018hat}
P.~Braun-Munzinger and B.~D\"onigus, ``{Loosely-bound objects produced in
  nuclear collisions at the LHC}'',
  \href{http://dx.doi.org/10.1016/j.nuclphysa.2019.02.006}{{\em Nucl. Phys. A}
  {\bfseries 987} (2019) 144--201},
  \href{http://arxiv.org/abs/1809.04681}{{\ttfamily arXiv:1809.04681
  [nucl-ex]}}.

\bibitem{Steinheimer:2012tb}
J.~Steinheimer, K.~Gudima, A.~Botvina, I.~Mishustin, M.~Bleicher, and
  H.~Stocker, ``{Hypernuclei, dibaryon and antinuclei production in high energy
  heavy ion collisions: Thermal production versus Coalescence}'',
  \href{http://dx.doi.org/10.1016/j.physletb.2012.06.069}{{\em Phys. Lett. B}
  {\bfseries 714} (2012) 85--91},
  \href{http://arxiv.org/abs/1203.2547}{{\ttfamily arXiv:1203.2547 [nucl-th]}}.

\bibitem{Braun-Munzinger:1994zkz}
P.~Braun-Munzinger and J.~Stachel, ``{Production of strange clusters and
  strange matter in nucleus-nucleus collisions at the AGS}'',
  \href{http://dx.doi.org/10.1088/0954-3899/21/3/002}{{\em J. Phys. G}
  {\bfseries 21} (1995) L17--L20},
  \href{http://arxiv.org/abs/nucl-th/9412035}{{\ttfamily
  arXiv:nucl-th/9412035}}.

\bibitem{Andronic:2010qu}
A.~Andronic, P.~Braun-Munzinger, J.~Stachel, and H.~Stoecker, ``{Production of
  light nuclei, hypernuclei and their antiparticles in relativistic nuclear
  collisions}'', \href{http://dx.doi.org/10.1016/j.physletb.2011.01.053}{{\em
  Phys. Lett.} {\bfseries B697} (2011) 203--207},
  \href{http://arxiv.org/abs/1010.2995}{{\ttfamily arXiv:1010.2995 [nucl-th]}}.

\bibitem{Cleymans:2011pe}
J.~Cleymans, S.~Kabana, I.~Kraus, H.~Oeschler, K.~Redlich, {\em et~al.},
  ``{Antimatter production in proton-proton and heavy-ion collisions at
  ultrarelativistic energies}'',
  \href{http://dx.doi.org/10.1103/PhysRevC.84.054916}{{\em Phys. Rev.}
  {\bfseries C84} (2011) 054916},
\href{http://arxiv.org/abs/1105.3719}{{\ttfamily arXiv:1105.3719 [hep-ph]}}.
%%CITATION = ARXIV:1105.3719;%%.

\bibitem{Vovchenko:2018fiy}
V.~Vovchenko, B.~D{\"o}nigus, and H.~Stoecker, ``{Multiplicity dependence of
  light nuclei production at LHC energies in the canonical statistical
  model}'', \href{http://dx.doi.org/10.1016/j.physletb.2018.08.041}{{\em Phys.
  Lett.} {\bfseries B785} (2018) 171--174},
\href{http://arxiv.org/abs/1808.05245}{{\ttfamily arXiv:1808.05245 [hep-ph]}}.
%%CITATION = ARXIV:1808.05245;%%.

\bibitem{Butler:1963pp}
S.~Butler and C.~Pearson, ``{Deuterons from High-Energy Proton Bombardment of
  Matter}'',
\href{http://dx.doi.org/10.1103/PhysRev.129.836}{{\em Phys. Rev.} {\bfseries
  129} (1963) 836--842}.
%%CITATION = PHRVA,129,836;%%.

\bibitem{Scheibl:1998tk}
R.~Scheibl and U.~W. Heinz, ``{Coalescence and flow in ultrarelativistic heavy
  ion collisions}'', \href{http://dx.doi.org/10.1103/PhysRevC.59.1585}{{\em
  Phys. Rev.} {\bfseries C59} (1999) 1585--1602},
\href{http://arxiv.org/abs/nucl-th/9809092}{{\ttfamily arXiv:nucl-th/9809092
  [nucl-th]}}.
%%CITATION = NUCL-TH/9809092;%%.

\bibitem{Blum:2019suo}
K.~Blum and M.~Takimoto, ``{Nuclear coalescence from correlation functions}'',
  \href{http://dx.doi.org/10.1103/PhysRevC.99.044913}{{\em Phys. Rev. C}
  {\bfseries 99} (2019) 044913},
  \href{http://arxiv.org/abs/1901.07088}{{\ttfamily arXiv:1901.07088
  [nucl-th]}}.

\bibitem{Mrowczynski:2016xqm}
S.~Mrowczynski, ``{Production of light nuclei in the thermal and coalescence
  models}'', \href{http://dx.doi.org/10.5506/APhysPolB.48.707}{{\em Acta Phys.
  Polon. B} {\bfseries 48} (2017) 707},
  \href{http://arxiv.org/abs/1607.02267}{{\ttfamily arXiv:1607.02267
  [nucl-th]}}.

\bibitem{Mrowczynski:1994rn}
S.~Mrowczynski, ``{Sum rule of the correlation function}'',
  \href{http://dx.doi.org/10.1016/0370-2693(94)01631-L}{{\em Phys. Lett. B}
  {\bfseries 345} (1995) 393--396},
  \href{http://arxiv.org/abs/hep-ph/9502215}{{\ttfamily arXiv:hep-ph/9502215}}.

\bibitem{Boschini:2020jty}
M.~Boschini {\em et~al.}, ``{Inference of the Local Interstellar Spectra of
  Cosmic-Ray Nuclei Z~\ensuremath{\leq}~28 with the GalProp\textendash{}HelMod
  Framework}'', \href{http://dx.doi.org/10.3847/1538-4365/aba901}{{\em
  Astrophys. J. Suppl.} {\bfseries 250} (2020) 27},
  \href{http://arxiv.org/abs/2006.01337}{{\ttfamily arXiv:2006.01337
  [astro-ph.HE]}}.

\bibitem{Abbas:2013rua}
{\bfseries ALICE} Collaboration, E.~Abbas {\em et~al.}, ``{Mid-rapidity
  anti-baryon to baryon ratios in pp collisions at $\sqrt{s}$ = 0.9, 2.76 and 7
  TeV measured by ALICE}'',
  \href{http://dx.doi.org/10.1140/epjc/s10052-013-2496-5}{{\em Eur. Phys. J. C}
  {\bfseries 73} (2013) 2496}, \href{http://arxiv.org/abs/1305.1562}{{\ttfamily
  arXiv:1305.1562 [nucl-ex]}}.

\bibitem{Acharya:2020cee}
{\bfseries ALICE} Collaboration, S.~Acharya {\em et~al.}, ``{Measurement of the
  low-energy antideuteron inelastic cross section}'',
  \href{http://dx.doi.org/10.1103/PhysRevLett.125.162001}{{\em Phys. Rev.
  Lett.} {\bfseries 125} (2020) 162001},
  \href{http://arxiv.org/abs/2005.11122}{{\ttfamily arXiv:2005.11122
  [nucl-ex]}}.

\bibitem{Aamodt:2008zz}
{\bfseries ALICE} Collaboration, K.~Aamodt {\em et~al.}, ``{The ALICE
  experiment at the CERN LHC}'',
  \href{http://dx.doi.org/10.1088/1748-0221/3/08/S08002}{{\em JINST} {\bfseries
  3} (2008) S08002}.

\bibitem{Abelev:2014ffa}
{\bfseries ALICE} Collaboration, B.~Abelev {\em et~al.}, ``{Performance of the
  ALICE Experiment at the CERN LHC}'',
  \href{http://dx.doi.org/10.1142/S0217751X14300440}{{\em Int. J. Mod. Phys.}
  {\bfseries A29} (2014) 1430044},
\href{http://arxiv.org/abs/1402.4476}{{\ttfamily arXiv:1402.4476 [nucl-ex]}}.
%%CITATION = ARXIV:1402.4476;%%.

\bibitem{Aamodt:2010dx}
{\bfseries ALICE} Collaboration, K.~Aamodt {\em et~al.}, ``{Midrapidity
  antiproton-to-proton ratio in pp collisions at $\sqrt{s} = 0.9$ and $7$~TeV
  measured by the ALICE experiment}'',
  \href{http://dx.doi.org/10.1103/PhysRevLett.105.072002}{{\em Phys. Rev.
  Lett.} {\bfseries 105} (2010) 072002},
\href{http://arxiv.org/abs/1006.5432}{{\ttfamily arXiv:1006.5432 [hep-ex]}}.
%%CITATION = ARXIV:1006.5432;%%.

\bibitem{TRDNote}
{\bfseries ALICE} Collaboration, ``{Validation of the ALICE material budget
  between TPC and TOF detectors}'', {\em ALICE-PUBLIC-2022-001} (Feb, 2022) .
  \url{https://cds.cern.ch/record/2800896}.

\bibitem{Agostinelli:2002hh}
{\bfseries \geant} Collaboration, S.~Agostinelli {\em et~al.}, ``{\geant: A
  Simulation toolkit}'',
\href{http://dx.doi.org/10.1016/S0168-9002(03)01368-8}{{\em Nucl.Instrum.Meth.}
  {\bfseries A506} (2003) 250--303}.
%%CITATION = NUIMA,A506,250;%%.

\bibitem{Uzhinsky:2011zz}
V.~Uzhinsky, J.~Apostolakis, A.~Galoyan, G.~Folger, V.~Grichine, {\em et~al.},
  ``{Antinucleus-nucleus cross sections implemented in \geant}'',
\href{http://dx.doi.org/10.1016/j.physletb.2011.10.010}{{\em Phys. Lett.}
  {\bfseries B705} (2011) 235--239}.
%%CITATION = PHLTA,B705,235;%%.

\bibitem{Lambert}
J.~H. Lambert, {\em {Photometria, sive De mensura et gradibus luminis, colorum
  et umbrae.}}
\newblock 1760.

\bibitem{Cudell:2001pn}
J.~Cudell, V.~Ezhela, P.~Gauron, K.~Kang, Y.~Kuyanov, S.~Lugovsky,
  B.~Nicolescu, and N.~Tkachenko, ``{Hadronic scattering amplitudes:
  Medium-energy constraints on asymptotic behavior}'',
  \href{http://dx.doi.org/10.1103/PhysRevD.65.074024}{{\em Phys. Rev. D}
  {\bfseries 65} (2002) 074024},
  \href{http://arxiv.org/abs/hep-ph/0107219}{{\ttfamily arXiv:hep-ph/0107219}}.

\bibitem{Shukla:2020bql}
A.~Shukla, A.~Datta, P.~von Doetinchem, D.-M. Gomez-Coral, and C.~Kanitz,
  ``{Large-scale Simulations of Antihelium Production in Cosmic-ray
  Interactions}'', \href{http://dx.doi.org/10.1103/PhysRevD.102.063004}{{\em
  Phys. Rev. D} {\bfseries 102} (2020) 063004},
  \href{http://arxiv.org/abs/2006.12707}{{\ttfamily arXiv:2006.12707
  [astro-ph.HE]}}.

\bibitem{Tomassetti:2017izg}
N.~Tomassetti and A.~Oliva, ``{Production and acceleration of antinuclei in
  supernova shockwaves}'',
  \href{http://dx.doi.org/10.3847/2041-8213/aa80da}{{\em Astrophys. J. Lett.}
  {\bfseries 844} (2017) L26},
  \href{http://arxiv.org/abs/1707.06915}{{\ttfamily arXiv:1707.06915
  [astro-ph.HE]}}.

\bibitem{Herms:2016vop}
J.~Herms, A.~Ibarra, A.~Vittino, and S.~Wild, ``{Antideuterons in cosmic rays:
  sources and discovery potential}'',
  \href{http://dx.doi.org/10.1088/1475-7516/2017/02/018}{{\em JCAP} {\bfseries
  02} (2017) 018}, \href{http://arxiv.org/abs/1610.00699}{{\ttfamily
  arXiv:1610.00699 [astro-ph.HE]}}.

\bibitem{Barrau:2002mc}
A.~Barrau, G.~Boudoul, F.~Donato, D.~Maurin, P.~Salati, I.~Stefanon, and
  R.~Taillet, ``{Antideuterons as a probe of primordial black holes}'',
  \href{http://dx.doi.org/10.1051/0004-6361:20021588}{{\em Astron. Astrophys.}
  {\bfseries 398} (2003) 403--410},
  \href{http://arxiv.org/abs/astro-ph/0207395}{{\ttfamily
  arXiv:astro-ph/0207395}}.

\bibitem{Serksnyte:2022onw}
L.~\v{S}erk\v{s}nyt\.{e} {\em et~al.}, ``{Reevaluation of the cosmic
  antideuteron flux from cosmic-ray interactions and from exotic sources}'',
  \href{http://dx.doi.org/10.1103/PhysRevD.105.083021}{{\em Phys. Rev. D}
  {\bfseries 105} (2022) 083021},
  \href{http://arxiv.org/abs/2201.00925}{{\ttfamily arXiv:2201.00925
  [astro-ph.HE]}}.

\bibitem{Navarro:1995iw}
J.~F. Navarro, C.~S. Frenk, and S.~D. White, ``{The Structure of cold dark
  matter halos}'', \href{http://dx.doi.org/10.1086/177173}{{\em Astrophys. J.}
  {\bfseries 462} (1996) 563--575},
  \href{http://arxiv.org/abs/astro-ph/9508025}{{\ttfamily
  arXiv:astro-ph/9508025}}.

\bibitem{Kissmann:2017}
R.~Kissmann, ``Galactic cosmic ray propagation models using picard'',
  \href{http://dx.doi.org/10.1088/1742-6596/837/1/012003}{{\em Journal of
  Physics: Conference Series} {\bfseries 837} (May, 2017) 012003}.
  \url{https://doi.org/10.1088/1742-6596/837/1/012003}.

\bibitem{Kissmann:2014sia}
R.~Kissmann, ``{PICARD: A novel code for the Galactic Cosmic Ray propagation
  problem}'', \href{http://dx.doi.org/10.1016/j.astropartphys.2014.02.002}{{\em
  Astropart. Phys.} {\bfseries 55} (2014) 37--50},
  \href{http://arxiv.org/abs/1401.4035}{{\ttfamily arXiv:1401.4035
  [astro-ph.HE]}}.

\bibitem{Evoli:2008}
C.~Evoli, D.~Gaggero, D.~Grasso, and L.~Maccione, ``Cosmic ray nuclei,
  antiprotons and gamma rays in the galaxy: a new diffusion model'',
  \href{http://dx.doi.org/10.1088/1475-7516/2008/10/018}{{\em Journal of
  Cosmology and Astroparticle Physics} {\bfseries 2008} (Oct, 2008) 018}.
  \url{https://doi.org/10.1088/1475-7516/2008/10/018}.

\bibitem{Strong:1998pw}
A.~Strong and I.~Moskalenko, ``{Propagation of cosmic-ray nucleons in the
  galaxy}'', \href{http://dx.doi.org/10.1086/306470}{{\em Astrophys. J.}
  {\bfseries 509} (1998) 212--228},
  \href{http://arxiv.org/abs/astro-ph/9807150}{{\ttfamily
  arXiv:astro-ph/9807150}}.

\bibitem{Moskalenko:2001ya}
I.~V. Moskalenko, A.~W. Strong, J.~F. Ormes, and M.~S. Potgieter, ``{Secondary
  anti-protons and propagation of cosmic rays in the galaxy and heliosphere}'',
  \href{http://dx.doi.org/10.1086/324402}{{\em Astrophys. J.} {\bfseries 565}
  (2002) 280--296}, \href{http://arxiv.org/abs/astro-ph/0106567}{{\ttfamily
  arXiv:astro-ph/0106567}}.

\bibitem{Bronfman:1988}
L.~{Bronfman}, R.~S. {Cohen}, H.~{Alvarez}, J.~{May}, and P.~{Thaddeus}, ``{A
  CO Survey of the Southern Milky Way: The Mean Radial Distribution of
  Molecular Clouds within the Solar Circle}'',
  \href{http://dx.doi.org/10.1086/165892}{{\em Astrophysical Journal}
  {\bfseries 324} (Jan., 1988) 248}.

\bibitem{Gordon:1976}
M.~A. {Gordon} and W.~B. {Burton}, ``{Carbon monoxide in the Galaxy. I. The
  radial distribution of CO, H$_{2}$, and nucleons.}'',
  \href{http://dx.doi.org/10.1086/154613}{{\em Astrophysical Journal}
  {\bfseries 208} (Sept., 1976) 346--353}.

\bibitem{Cordes:1991}
J.~M. {Cordes}, J.~M. {Weisberg}, D.~A. {Frail}, S.~R. {Spangler}, and
  M.~{Ryan}, ``{The galactic distribution of free electrons}'',
  \href{http://dx.doi.org/10.1038/354121a0}{{\em Nature} {\bfseries 354} (Nov.,
  1991) 121--124}.

\bibitem{Dickey:1990}
J.~M. {Dickey} and F.~J. {Lockman}, ``{H I in the galaxy.}'',
  \href{http://dx.doi.org/10.1146/annurev.aa.28.090190.001243}{{\em Annual
  Review of Astronomy and Astrophysics} {\bfseries 28} (Jan., 1990) 215--261}.

\bibitem{Gleeson:1968zza}
L.~Gleeson and W.~Axford, ``{Solar Modulation of Galactic Cosmic Rays}'',
  \href{http://dx.doi.org/10.1086/149822}{{\em Astrophys. J.} {\bfseries 154}
  (1968) 1011}.

\bibitem{Boschini:2019ubh}
M.~J. Boschini, S.~Della~Torre, M.~Gervasi, G.~La~Vacca, and P.~G. Rancoita,
  ``{The HelMod Model in the Works for Inner and Outer Heliosphere: from AMS to
  Voyager Probes Observations}'',
  \href{http://dx.doi.org/10.1016/j.asr.2019.04.007}{{\em Adv. Space Res.}
  {\bfseries 64} (2019) 2459--2476},
  \href{http://arxiv.org/abs/1903.07501}{{\ttfamily arXiv:1903.07501
  [physics.space-ph]}}.

\bibitem{Cummings:2016pdr}
A.~Cummings, E.~Stone, B.~Heikkila, N.~Lal, W.~Webber, G.~J\'ohannesson,
  I.~Moskalenko, E.~Orlando, and T.~Porter, ``{Galactic Cosmic Rays in the
  Local Interstellar Medium: Voyager 1 Observations and Model Results}'',
  \href{http://dx.doi.org/10.3847/0004-637X/831/1/18}{{\em Astrophys. J.}
  {\bfseries 831} (2016) 18}.

\bibitem{Aguilar:2015ooa}
{\bfseries AMS} Collaboration, M.~Aguilar {\em et~al.}, ``{Precision
  Measurement of the Proton Flux in Primary Cosmic Rays from Rigidity 1 GV to
  1.8 TV with the Alpha Magnetic Spectrometer on the International Space
  Station}'', \href{http://dx.doi.org/10.1103/PhysRevLett.114.171103}{{\em
  Phys. Rev. Lett.} {\bfseries 114} (2015) 171103}.

\bibitem{Engelmann:1990zz}
J.~Engelmann, P.~Ferrando, A.~Soutoul, P.~Goret, and E.~Juliusson, ``{Charge
  composition and energy spectra of cosmic-ray for elements from Be to NI -
  Results from HEAO-3-C2}'', {\em Astron. Astrophys.} {\bfseries 233} (1990)
  96--111.

\bibitem{Ahn:2008my}
H.~Ahn {\em et~al.}, ``{Measurements of cosmic-ray secondary nuclei at high
  energies with the first flight of the CREAM balloon-borne experiment}'',
  \href{http://dx.doi.org/10.1016/j.astropartphys.2008.07.010}{{\em Astropart.
  Phys.} {\bfseries 30} (2008) 133--141},
  \href{http://arxiv.org/abs/0808.1718}{{\ttfamily arXiv:0808.1718
  [astro-ph]}}.

\bibitem{GAPS:2020axg}
{\bfseries GAPS} Collaboration, N.~Saffold {\em et~al.}, ``{Cosmic antihelium-3
  nuclei sensitivity of the GAPS experiment}'',
  \href{http://dx.doi.org/10.1016/j.astropartphys.2021.102580}{{\em Astropart.
  Phys.} {\bfseries 130} (2021) 102580},
  \href{http://arxiv.org/abs/2012.05834}{{\ttfamily arXiv:2012.05834
  [hep-ph]}}.

\bibitem{Cuoco:2016eej}
A.~Cuoco, M.~Kr\"amer, and M.~Korsmeier, ``{Novel Dark Matter Constraints from
  Antiprotons in Light of AMS-02}'',
  \href{http://dx.doi.org/10.1103/PhysRevLett.118.191102}{{\em Phys. Rev.
  Lett.} {\bfseries 118} (2017) 191102},
  \href{http://arxiv.org/abs/1610.03071}{{\ttfamily arXiv:1610.03071
  [astro-ph.HE]}}.

\bibitem{Adam:2015pza}
{\bfseries ALICE} Collaboration, J.~Adam {\em et~al.}, ``{Pseudorapidity and
  transverse-momentum distributions of charged particles in
  proton\textendash{}proton collisions at $\sqrt s=$ 13 TeV}'',
  \href{http://dx.doi.org/10.1016/j.physletb.2015.12.030}{{\em Phys. Lett. B}
  {\bfseries 753} (2016) 319--329},
  \href{http://arxiv.org/abs/1509.08734}{{\ttfamily arXiv:1509.08734
  [nucl-ex]}}.

\bibitem{Adam:2015ptt}
{\bfseries ALICE} Collaboration, J.~Adam {\em et~al.}, ``{Centrality dependence
  of the charged-particle multiplicity density at midrapidity in Pb-Pb
  collisions at $\sqrt{s_{\rm NN}}$ = 5.02 TeV}'',
  \href{http://dx.doi.org/10.1103/PhysRevLett.116.222302}{{\em Phys. Rev.
  Lett.} {\bfseries 116} (2016) 222302},
  \href{http://arxiv.org/abs/1512.06104}{{\ttfamily arXiv:1512.06104
  [nucl-ex]}}.

\bibitem{PRLsupplemental}
 See Supplemental Material for cumulative distribution of the material in the
  ALICE apparatus at
  http://link.aps.org/supplemental/10.1103/PhysRevLett.125.162001.

\bibitem{Wang:1991hta}
X.-N. Wang and M.~Gyulassy, ``{HIJING: A Monte Carlo model for multiple jet
  production in p p, p A and A A collisions}'',
  \href{http://dx.doi.org/10.1103/PhysRevD.44.3501}{{\em Phys. Rev. D}
  {\bfseries 44} (1991) 3501--3516}.

\bibitem{ALICE:2005vhb}
{\bfseries ALICE} Collaboration, C.~W. Fabjan {\em et~al.}, ``{ALICE: Physics
  performance report, volume II}'',
  \href{http://dx.doi.org/10.1088/0954-3899/32/10/001}{{\em J. Phys. G}
  {\bfseries 32} (2006) 1295--2040}.

\bibitem{ALICE:2010mlf}
{\bfseries ALICE} Collaboration, K.~Aamodt {\em et~al.}, ``{Centrality
  dependence of the charged-particle multiplicity density at mid-rapidity in
  Pb-Pb collisions at $\sqrt{s_{NN}}=2.76$ TeV}'',
  \href{http://dx.doi.org/10.1103/PhysRevLett.106.032301}{{\em Phys. Rev.
  Lett.} {\bfseries 106} (2011) 032301},
  \href{http://arxiv.org/abs/1012.1657}{{\ttfamily arXiv:1012.1657 [nucl-ex]}}.

\bibitem{Zyla:2020zbs}
{\bfseries Particle Data Group} Collaboration, P.~A. Zyla {\em et~al.},
  ``{Review of Particle Physics}'',
  \href{http://dx.doi.org/10.1093/ptep/ptaa104}{{\em PTEP} {\bfseries 2020}
  (2020) 083C01}.

\bibitem{Ingemarsson:2001uil}
A.~Ingemarsson {\em et~al.}, ``{Reaction cross sections of intermediate energy
  3 He-particles on targets from 9 Be to 208 Pb}'',
  \href{http://dx.doi.org/10.1016/S0375-9474(01)01116-2}{{\em Nucl. Phys. A}
  {\bfseries 696} (2001) 3--30}.

\bibitem{Lin:2019yux}
H.-N. Lin and X.~Li, ``{The Dark Matter Profiles in the Milky Way}'',
  \href{http://dx.doi.org/10.1093/mnras/stz1698}{{\em Mon. Not. Roy. Astron.
  Soc.} {\bfseries 487} (2019) 5679--5684},
  \href{http://arxiv.org/abs/1906.08419}{{\ttfamily arXiv:1906.08419
  [astro-ph.GA]}}.

\bibitem{PropagationNote}
{\bfseries ALICE} Collaboration, ``{Modelling of Antihelium-3 Cosmic-Ray
  Propagation}'', {\em ALICE-PUBLIC-2022-002} (Feb, 2022) .
  \url{https://cds.cern.ch/record/2800897}.

\bibitem{Duperray:2005si}
R.~Duperray, B.~Baret, D.~Maurin, G.~Boudoul, A.~Barrau, L.~Derome,
  K.~Protasov, and M.~Buenerd, ``{Flux of light antimatter nuclei near Earth,
  induced by cosmic rays in the Galaxy and in the atmosphere}'',
  \href{http://dx.doi.org/10.1103/PhysRevD.71.083013}{{\em Phys. Rev. D}
  {\bfseries 71} (2005) 083013},
  \href{http://arxiv.org/abs/astro-ph/0503544}{{\ttfamily
  arXiv:astro-ph/0503544}}.

\bibitem{Gomez-Coral:2018yuk}
D.-M. Gomez-Coral, A.~Menchaca~Rocha, V.~Grabski, A.~Datta, P.~von Doetinchem,
  and A.~Shukla, ``{Deuteron and Antideuteron Production Simulation in
  Cosmic-Ray Interactions}'',
  \href{http://dx.doi.org/10.1103/PhysRevD.98.023012}{{\em Phys. Rev. D}
  {\bfseries 98} (2018) 023012},
  \href{http://arxiv.org/abs/1806.09303}{{\ttfamily arXiv:1806.09303
  [astro-ph.HE]}}.

\bibitem{Acharya:2017fvb}
{\bfseries ALICE} Collaboration, S.~Acharya {\em et~al.}, ``{Production of
  deuterons, tritons, $^{3}$He nuclei and their antinuclei in pp collisions at
  $\mathbf{\sqrt{{\textit s}}}$ = 0.9, 2.76 and 7 TeV}'',
  \href{http://dx.doi.org/10.1103/PhysRevC.97.024615}{{\em Phys. Rev.}
  {\bfseries C97} (2018) 024615},
\href{http://arxiv.org/abs/1709.08522}{{\ttfamily arXiv:1709.08522 [nucl-ex]}}.
%%CITATION = ARXIV:1709.08522;%%.

\bibitem{Acharya:2020sfy}
{\bfseries ALICE} Collaboration, S.~Acharya {\em et~al.}, ``{(Anti-)deuteron
  production in pp collisions at $\sqrt{s}=13 \ \text {TeV}$}'',
  \href{http://dx.doi.org/10.1140/epjc/s10052-020-8256-4}{{\em Eur. Phys. J. C}
  {\bfseries 80} (2020) 889}, \href{http://arxiv.org/abs/2003.03184}{{\ttfamily
  arXiv:2003.03184 [nucl-ex]}}.

\bibitem{Sjostrand:2007gs}
T.~Sjostrand, S.~Mrenna, and P.~Z. Skands, ``{A Brief Introduction to PYTHIA
  8.1}'', \href{http://dx.doi.org/10.1016/j.cpc.2008.01.036}{{\em Comput. Phys.
  Commun.} {\bfseries 178} (2008) 852--867},
  \href{http://arxiv.org/abs/0710.3820}{{\ttfamily arXiv:0710.3820 [hep-ph]}}.

\bibitem{Burkert:1995yz}
A.~Burkert, ``{The Structure of dark matter halos in dwarf galaxies}'',
  \href{http://dx.doi.org/10.1086/309560}{{\em Astrophys. J. Lett.} {\bfseries
  447} (1995) L25}, \href{http://arxiv.org/abs/astro-ph/9504041}{{\ttfamily
  arXiv:astro-ph/9504041}}.

\bibitem{Begeman:1991iy}
K.~G. Begeman, A.~H. Broeils, and R.~H. Sanders, ``{Extended rotation curves of
  spiral galaxies: Dark haloes and modified dynamics}'', {\em Mon. Not. Roy.
  Astron. Soc.} {\bfseries 249} (1991) 523.

\end{thebibliography}\endgroup

\section{Methods}
\label{sec:Methods}
\subsection{Event selection}
The inelastic pp and Pb--Pb events were recorded with the ALICE apparatus at collision energies of $\sqrt{s} = 13$ TeV and $\sqrt{s_{\rm NN}} = 5.02$ TeV, respectively. Events are triggered by the V0 detector comprising two plastic scintillator arrays placed on both sides of the interaction point and covering the pseudorapidity intervals of $2.8 < \eta < 5.1$ and $-3.7 < \eta < -1.7$. The pseudorapidity is defined as \mbox{$\eta = -\ln[\tan(\frac{\Theta}{2})]$}, where $\Theta$ is the polar angle of the particle with respect to the beam axis. The trigger condition is defined by the coincidence of signals in both arrays of the V0 detector. Together with two innermost layers of the ITS detector, the V0 is also used to reject background events like beam--gas interactions or collisions with mechanical structures of the beam line. For the analysis of pp data, a high-multiplicity trigger is employed to select only events with the total signal amplitude measured in the V0 detector above a certain threshold, which leads to a selection of about $0.17\%$ of the inelastic pp collisions with the highest V0 signal. In these events, the number of charged particles produced at midrapidity $|\eta| < 0.5$ is about six times higher than $\langle \dNdy \rangle = 5.31 \pm 0.18$ measured in inelastic pp collision at $\sqrt{s} = 13$ TeV~\cite{Adam:2015pza}. This facilitates the analysis of rarely produced (anti)\he\ nuclei. As for the Pb--Pb experimental data, 10\% of all inelastic events with the highest signal amplitude in the V0 detector are considered for the analysis. In these events, the average charged-particle multiplicity at midrapidity $|\eta| < 0.5$ amounts to $\langle \dNdy \rangle = 1764 \pm 50$~\cite{Adam:2015ptt}. In total, $147.9\times10^{6}$ Pb--Pb and $10^{9}$ pp events were analysed.

\subsection{Particle tracking and identification}

Trajectories of charged particles are reconstructed in the ALICE central barrel from their hits in the ITS  and TPC. The detectors are located inside a solenoidal magnetic field (0.5 T) bending the trajectories of charged particles. The curvature and direction of the charged-particle trajectories in the magnetic field are used to reconstruct their momentum. The detectors  provide full azimuthal coverage in the pseudorapidity interval $|\eta| < 0.9$. This $\eta$ range corresponds to the region within $\pm$42 degrees of the transverse plane that is perpendicular to the beam axis. Typical resolution of the transverse momentum reconstructed at the primary vertex ($p_{\rm{T, primary}}$) for protons, pions and kaons varies from about $2\%$ for tracks with $p_{\rm{T, primary}} =$ 10 GeV/$c$ to below $1\%$ for $p_{\rm{T, primary}}\leq$ 1 GeV/$c$.

Specific energy loss in the TPC gas is used to identify charged particles. Due to their electric charge ($z = 2$), high mass and quadratic dependence of specific energy loss on particle charge, \he\ and \ahe\ nuclei have larger energy loss than most other (anti)particles produced in the collision (like pions, kaons, protons and deuterons) and can be clearly identified in the TPC. 
The selected \he\ candidates include a substantial amount of background from secondary nuclei that originate from spallation reactions in the detector material and can be seen at low momentum (Fig.~\ref{fig:Scheme1}b). This contribution is estimated via a fit to the distribution of the measured distance of closest approach between the track candidates and the primary collision vertex using templates from MC simulations. Since primary particles point back to the primary vertex, they are characterized by a distinct peak structure at zero distance of closest approach, whereas secondary particles correspond to a flat distribution of the distance of closest approach and their contribution can, therefore, be separated. More details on this procedure can be found elsewhere~\cite{ALICE:2015wav,Abbas:2013rua}. For \he\ candidates in pp collisions at $\sqrt{s} = 13$ TeV this contribution amounts to $\sim 75\%$ in the lowest analysed momentum interval of $0.65 \leq p_{\rm primary}/z < 0.8$ GeV/$c$ and is negligible in the momentum range above $p_{\rm primary}/z = 1.5$ GeV/$c$. For \ahe\ nuclei, there is no contribution from spallation processes. In total, there are $16801 \pm 130$ primary \ahe\ reconstructed in the TPC in the Pb--Pb data sample. In the sample of pp collisions, the total number of reconstructed primary candidates of \he\ and \ahe\ is $773 \pm 46$ and $652 \pm 30$, respectively. The uncertainties for these values result from the fit to the TPC signal which is used to reject the (small) background from (anti)triton nuclei misidentified as (anti)\he\ at low momenta.

\subsection{Corrections and evaluation of the systematic uncertainties}
\label{methods:corrections}
Due to continuous energy-loss effects in the detector material, the inelastic interaction of \ahe\ with the detector material happens at momentum $p$, which is lower than the momentum $p_{\rm primary}$ reconstructed at the primary collision vertex. The corresponding effect is taken into account utilising MC simulations in which one has precise information about both momenta for each (anti)particle. In the analysis of pp collisions, the average values of $p / p_{\rm primary}$ distributions in each analysed $p_{\rm primary}$ interval are used to consider the energy loss. The root mean square (RMS) value of these distributions is used to determine the uncertainty of the momentum $p$, which is propagated to the uncertainty of the measured cross section. For the analysis of Pb--Pb data sample, the MC information on the momenta of daughter tracks originating from the \ahe\ annihilation is used to estimate the corresponding effect and resulting uncertainty.

The systematic uncertainties due to tracking, particle identification and the description of material budget in MC simulations are considered, and the total uncertainty is obtained as the quadratic sum of the individual contributions. The material budget of the ALICE apparatus~\cite{PRLsupplemental,Abelev:2014ffa,TRDNote} is varied by $\pm 4.5\%$ in MC simulations, and the deviations of the final results from the default case are considered as an uncertainty. The precision of $\sim 4.5\%$ of the MC parameterization is validated for the ALICE material with photon conversion analyses (up to the outer TPC vessel~\cite{Abelev:2014ffa}) and with tagged pion and proton absorption studies (for the material between TPC and TOF detectors~\cite{TRDNote}).

For the Pb--Pb analysis, the total systematic uncertainty amounts to $\sim 20\%$ in the highest and lowest momentum intervals considered in the analysis and decreases to $\leq 10\%$ in the momentum interval of $3\leq p<7$ GeV/$c$. For the analysis of pp data (which is based on the antibaryon-to-baryon ratio method), an additional uncertainty due to primordial antibaryon-to-baryon ratio produced in collisions is considered as a global uncertainty. The primordial antiproton-to-proton ratio of $0.998 \pm 0.015$ is extrapolated for the $\sqrt{s} = 13$ TeV collision energy from available measurements~\cite{Abbas:2013rua,Aamodt:2010dx}; furthermore, under the assumption that the (anti)\he\ yield is proportional to the cube of (anti)proton yield~\cite{Scheibl:1998tk}, the primary $^{3}\overline{\rm{He}}/^{3}\rm{He}$ ratio amounts to $0.994\pm 0.045$. This uncertainty is the dominant contribution to the total systematic uncertainty for the pp analysis, which amounts to $\sim 8\%$.

\subsection{MC Simulation}

The results presented in this paper are compared with the detailed MC simulations of the ALICE detector. The simulations start with the generation of (anti)particles at the primary collision vertex and the production of raw detector information, also taking into account inactive subdetector channels. The same reconstruction algorithms applied to real experimental data are employed to analyse the raw simulated data. For the pp analysis based on antimatter-to-matter ratio, the primordial $^{3}\overline{\rm{He}}/^{3}\rm{He}$ ratio of $0.994$ is used as an input for the MC simulations. Since the average multiplicity in pp collisions at midrapidity is low, no underlying event was simulated in this case. For the TOF-to-TPC analysis in Pb--Pb collisions, the simulations contain an underlying Pb--Pb event that was generated with the help of the HIJING event generator~\cite{Wang:1991hta,ALICE:2005vhb,ALICE:2010mlf}. On top of this underlying event, 160 \ahe\ nuclei were injected following the momentum distribution obtained from independent studies on \ahe\ production~\cite{ALICE:2015wav}.

For the propagation of (anti)particles through the detector material, the simulations rely on the \geant software package~\cite{Agostinelli:2002hh}, in which the inelastic cross section of \ahe\ nuclei is based on Glauber calculations. Since the Glauber model simulations are computationally too expensive to be performed during the propagation steps through the material, they are parameterized as a function of the atomic mass number $A$ of the target nucleus~\cite{Uzhinsky:2011zz}:

\begin{equation}
    \sigma^{\rm inel}_{hA} = \pi R_{A}^{2} \ln\left(1+\frac{A\sigma^{\rm tot}_{hN}}{\pi R^{2}_{A}}\right).
\end{equation}

Here $h$ denotes the nucleus in question ($h$ = $\mathrm{\overline{p}}$, $\mathrm{\overline{d}}$, $^3\mathrm{\overline{\rm He}}$ and $^4\mathrm{\overline{\rm He}}$), and $A$ is the atomic number of the target nucleus with radius $R_{A}$. Also, $\sigma^{\rm tot}_{hN}$ is the total (elastic plus inelastic) cross section of hadron $h$ on nucleon $N$, which is estimated with the help of Glauber calculations by extrapolating the measured $\overline{\rm p}$p values~\cite{Cudell:2001pn} to larger antinuclei. We performed several full-scale MC simulations with varied inelastic cross sections of \ahe\ with matter, and the simulated observables used in this analysis are studied as a function of the inelastic cross section re-scaling. This dependence is parameterized using the Lambert--Beer law (Fig.~\ref{fig:Scheme1}e,f). The parameterization reads as $ N_{\rm surv} = N_0 \times \mathrm{exp} (-\sigma_{\rm inel} \rho L)$, where $N_0$ corresponds to the number of incident particles, $N_{\rm surv}$ is the number of survived particles that did not get absorbed, $\sigma_{inel}$ is the inelastic cross section, $\rho$ is the density of the material crossed, and $L$ is the length of the particle trajectory in the material. The free parameter given by the product $\rho L$ is determined by a fit to the simulated observables.

\begin{figure}[htbp]
\centering
\includegraphics[width=0.48\textwidth]{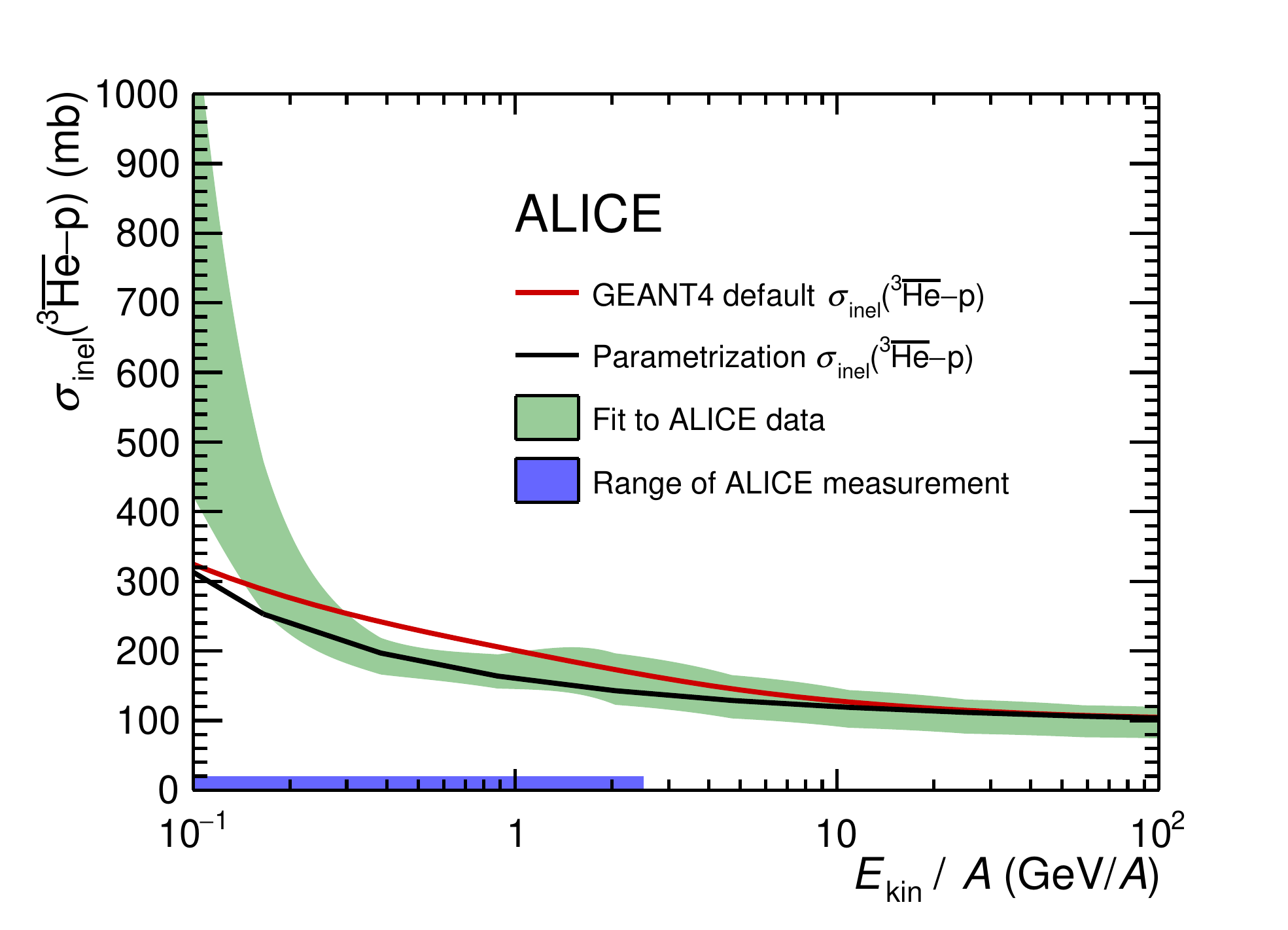}
\includegraphics[width=0.48\textwidth]{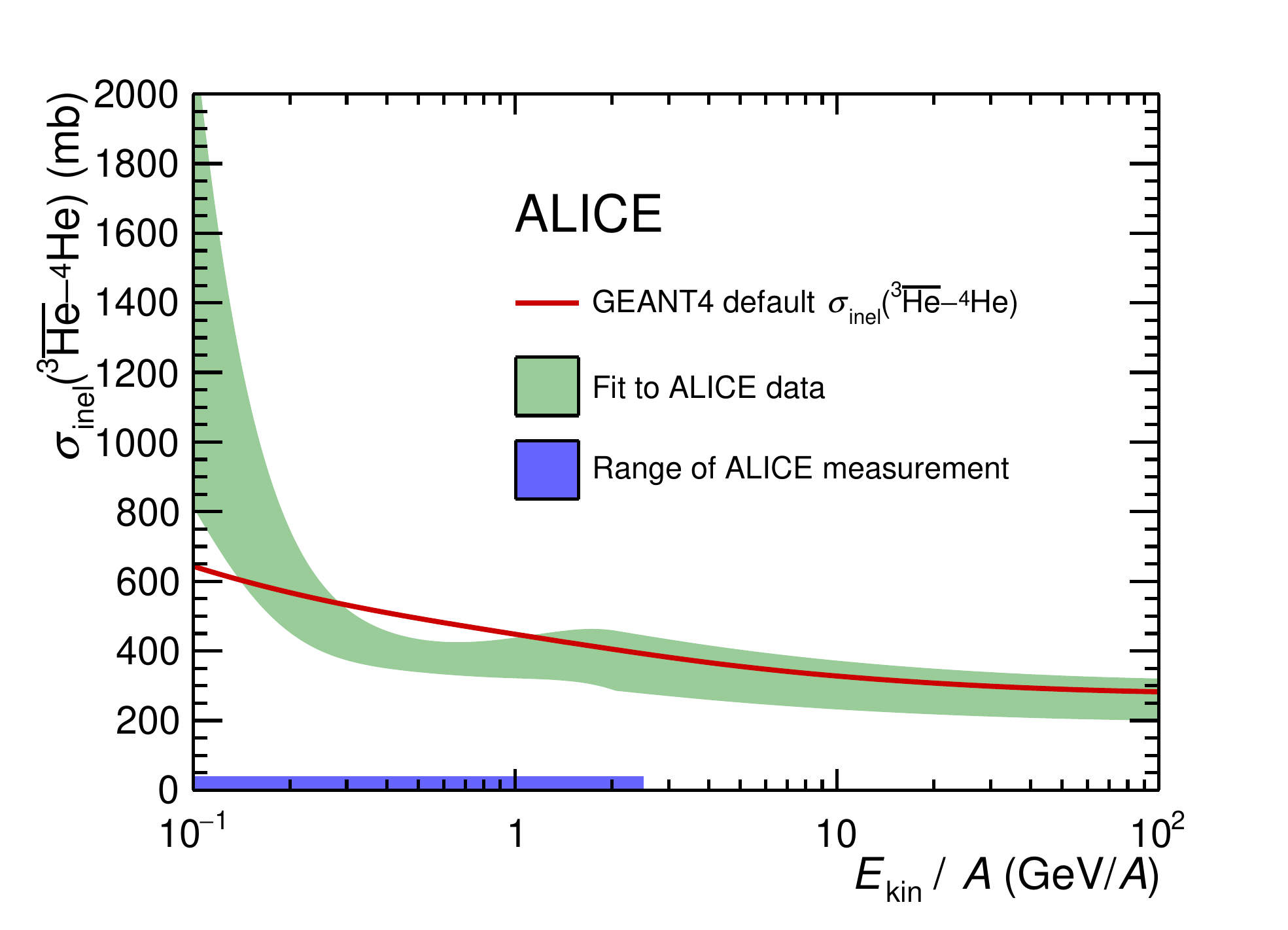}
\caption{Inelastic cross section for \ahe\ on protons (left) and on $^{4}$He (right). The green band shows the scaled ALICE measurement (see text for details), the red line represents the original \geant parameterization and the black line on the left plot the parameterization employed in Ref.~\cite{Korsmeier:2017xzj}. The width of the green band represents standard deviation uncertainty. The blue band on the x axis indicates the kinetic energy range corresponding to the ALICE measurement for \sigmainel.} \label{fig:CrossSectionScaled}
\end{figure}

To model the inelastic cross section of \ahe\ nuclei in the interstellar medium, the \geant parameterization of the \ahe--p inelastic cross section is scaled with the correction factors obtained from the ALICE measurements. The additional uncertainty that originates from re-scaling a measurement at $\langle A \rangle = 17.4$ and $\langle A \rangle = 34.7$ to $A = 1$ and $A = 4$ is taken from the difference between the parameterization for the dependence on $A$ in \geant and in full Glauber calculation and amounts to $<$ 8\%~\cite{Uzhinsky:2011zz}. The resulting \ahe--p inelastic cross section is shown in \nameref{sec:ExtendedData} Fig.~\ref{fig:CrossSectionScaled} (left) together with the model employed in another work~\cite{Korsmeier:2017xzj}. The latter is based on the approximation which uses available measurements to estimate the inelastic antideuteron--proton cross section in the following way:

\begin{equation}
    \sigma_{\rm inel}^{\overline{\rm d}\rm{p}} \approx \frac{\sigma_{\rm tot}^{\overline{\rm d}\rm{p}}}{\sigma_{\rm tot}^{\overline{\rm p}\rm{p}}}(\sigma_{\rm tot}^{\overline{\rm p}\rm{p}} - \sigma_{\rm el}^{\overline{\rm p}\rm{p}}).
\end{equation}

By symmetry the total antideuteron--proton cross section $\sigma_{\rm tot}^{\overline{\rm d}\rm{p}}$ is equal to the total deuteron--antiproton cross section taken from elsewhere~\cite{Zyla:2020zbs}. For antihelium, the inelastic cross section is scaled from antideuterons according to the mass number as $\sigma_{\rm inel}^{^{3}\overline{\rm He}\rm{p}} = \frac{3}{2} \sigma_{\rm inel}^{\overline{\rm d}\rm{p}}$. Figure~\ref{fig:CrossSectionScaled} (right) also shows the resulting \ahe--$^{4}$He inelastic cross section obtained in the same way for $^{4}$He target.

The results for the inelastic \ahe\ cross section are also tested against the modifications of elastic cross sections of \ahe\ nuclei. Both \he\ and \ahe\ elastic cross sections are independently varied by 30\%, which led to $\leq 1\%$ modifications of the final results. For the analysis of proton--proton collisions based on the antibaryon-to-baryon ratio method, the results are additionally investigated for the sensitivity to the \he\ inelastic cross section. The latter is varied by $10\%$, which is the uncertainty of the \geant parameterizations obtained from fits to experimental data~\cite{Ingemarsson:2001uil}. This variation yields a modification of $\leq 2.3\%$ in the reconstructed antihelium-to-helium ratio.

\subsection{Propagation modelling}
\label{methods:propagation}

The possible sources of antinuclei in our Galaxy are either cosmic-ray interactions with nuclei in the interstellar gas or more exotic sources such as DM annihilations or decays. Cosmic rays consist mainly of protons and originate from supernovae remnants, whereas DM has so far escaped direct or indirect detection but its density profile can be modelled~\cite{Lin:2019yux}.

The propagation in the Galaxy can be carried out using the publicly available propagation models~\cite{Kissmann:2017,Kissmann:2014sia,Evoli:2008,Strong:1998pw}. We choose the GALPROP code (version 56 available at https://galprop.stanford.edu) for the implementation of \ahe\ cosmic-ray propagation, which is discussed in details elsewhere~\cite{PropagationNote}. GALPROP numerically solves a general transport equation for all included particle species~\cite{Strong:1998pw}. This transport equation reads as

\begin{equation}
\label{eqn:TransportEquation}
\frac{\partial\psi}{\partial t} =q(\textbf{r},p) + \nabla \cdot (D_{xx}\textbf{grad}\psi-\textbf{V}\psi)+\frac{\partial}{\partial p}p^2D_{pp}\frac{\partial }{\partial p}\frac{\psi}{p^2}-\frac{\partial}{\partial p} \left[ \psi \frac{\mathrm{d}p}{\mathrm{d}t}-\frac{p}{3}(\nabla\cdot \textbf{V})\psi\right]-\frac{\psi}{\tau}.
\end{equation}
Here, $\psi = \psi(\textbf{r}, p, t)$ is the time-dependent \ahe\ density per unit of the total particle momentum and $q(\textbf{r},p) $ is the source function for \ahe\ . The second and third terms describe the propagation of \ahe\, where the $D_{xx}$, $\textbf{V}$ and $D_{pp}$ are the spatial diffusion coefficient, convection velocity and diffusive re-acceleration coefficient, respectively. Although the effect of the Galactic magnetic field is not explicitly modelled, it is accounted for by these terms of the transport equation. These coefficients are the same for all the particle species and can be constrained using available cosmic-ray measurements. We use the best-fit values of these parameters provided elsewhere~\cite{Boschini:2020jty}. The fourth term accounts for momentum losses via cosmic-ray interactions with interstellar gas ($\mathrm{d}p/\mathrm{d}t$) and adiabatic momentum losses ($\nabla\cdot \textbf{V}$). The last term represents the \ahe\ inelastic collisions with interstellar gas, where $1/\tau$ is the fragmentation rate. It is related to the inelastic cross section as follows:
\begin{equation}
\label{eqn:AnnihilationEquation}
\frac{1}{\tau}=\beta c\left(n_{H}(\textbf{r})\sigma_{\rm inel}^{^{3}\overline{\rm He}\rm{p}}(p) +n_{He}(\textbf{r})\sigma_{\rm inel}^{^{3}\overline{\rm He}^{4}\rm He}(p) \right).
\end{equation}
\\
The elastic re-scattering of cosmic-ray antinuclei in the interstellar medium (ISM) is assumed to have a negligible effect on diffusive propagation~\cite{Duperray:2005si}. The second and third term of Eq.~\ref{eqn:TransportEquation} can cause both acceleration and deceleration, which means that the final flux at a given energy also depends on the initial fluxes at both higher and lower energies. Therefore, the final number of particles in a specific energy interval depends on: (1) the energy spectrum and spatial distribution of the source, (2) the propagation parameters, (3) the particles’ momentum loss/gain and (4) the annihilation cross section. 
Only the first and last terms of Eq.~\ref{eqn:TransportEquation} require particle-specific information. Here \ahe\ nuclei can be produced when cosmic-ray particles interact with protons or $^4$He nuclei in ISM.
The \ahe\ source function in this case is
\begin{equation}
\label{eqn:SourceFromCR}
q(\textbf{r},p)=\sum\limits_{\mathrm{CR=H,He}} \; \sum\limits_{\mathrm{ISM=H,He}} n_{\mathrm{ISM}}(\textbf{r}) \int \mathrm{d} p^{\prime}_{\mathrm{CR}}\; \beta_{\mathrm{CR}} \; c  \frac{\mathrm{d} \sigma\left(p, p^{\prime}_{\mathrm{CR}}\right)}{\mathrm{d} p} \; n_{\mathrm{CR}}\left(\textbf{r},p^{\prime}_{\mathrm{CR}}\right).
\end{equation}
The density of hydrogen and helium gas is represented by $n_{\rm ISM}(\textbf{r}) $, and $p^{\prime}_{\rm CR}$, $\beta_{\rm CR}$ and $n_{\rm CR}(\textbf{r}, p'_{\rm CR})$ are the momentum, velocity and density of cosmic rays, respectively, whereas $p$ is the momentum of the produced \ahe. Also, $\mathrm{d}\sigma(p,p^{\prime}_{\rm CR})/\mathrm{d}p$ is the \ahe\ differential production cross section for the specific collision and includes primary \ahe\ as well as the products of $\overline{t}$ decays. The most abundant cosmic rays are protons and helium; thus, this source function must be calculated for both species and summed up. In another work~\cite{Shukla:2020bql}, all the relevant types of collisions between protons and $^{4}$He nuclei with projectile beam energies ranging from $31$ GeV to $12.5$ TeV are considered, and the so-called spherical approximation is used in which antinucleons with a momentum difference smaller than $p_{0}$ are forming an antinucleus~\cite{Shukla:2020bql,Gomez-Coral:2018yuk}. The parameter $p_{0}$ depends on collision energy and is constrained by several accelerator-based measurements~\cite{SimonGillo:1995dh,Armstrong:2000gd,Afanasev:2000ku,Anticic:2004yj,Adler:2004uy,Alper:1973my,Henning:1977mt,Alexopoulos:2000jk,Aktas:2004pq,Asner:2006pw,Schael:2006fd,ALICE:2015wav,Adam:2015yta,Acharya:2017dmc,Acharya:2019rgc,Agakishiev:2011ib,Abelev:2010}, including measurements at the LHC~\cite{Acharya:2017fvb,Acharya:2020sfy}. The resulting injection spectra obtained from the collisions of cosmic rays with the interstellar medium peak above 7 GeV/$A$~\cite{Shukla:2020bql}.

In the case of \ahe\ nuclei produced from DM annihilations, the source function depends on the thermally averaged inelastic cross section times the velocity ($\langle\sigma v\rangle$), density ($\rho_{\rm DM}$) of the DM, mass ($m_{\chi}$) of the DM particle and the resulting \ahe\ spectrum ($\mathrm{d}N/\mathrm{d}E_{\rm kin}$)~\cite{Carlson:2014ssa}:
\begin{equation}
\label{eqn:SourceFromDM}
q( \textbf{r},E_{\rm kin})=\frac{1}{2} \frac{\rho_{\mathrm{DM}}^{2}(\textbf{r})}{m_{\chi}^{2}}\langle\sigma v\rangle\frac{\mathrm{d} N}{\mathrm{d} E_{\rm kin}}.
\end{equation}
Here $E_{\rm kin}$ is the kinetic energy of the produced \ahe\ including those which are the products of $\overline{t}$ decays. The spectrum is calculated utilising the \textsc{Pythia} 8.156 event generator~\cite{Sjostrand:2007gs} and a coalescence model with a coalescence momentum $p_0 =$ 357 MeV/$c$, as described in more detail elsewhere~\cite{Carlson:2014ssa}.
We set $\langle\sigma v\rangle = 2.6 \times10^{-26}$ cm$^3$s$^{-1}$ (Ref.~\cite{Korsmeier:2017xzj}). We implemented the Navarro–Frenk–White profile in GALPROP, which is one of the most commonly used DM density profiles:
\begin{equation}
\label{eqn:NFWProfile}
\rho(r)=\frac{\rho_{0}}{\frac{r}{R_{s}}\left(1+\frac{r}{R_{s}}\right)^{2}}.
\end{equation}
Here $r$ is the distance to the Galactic Centre, $\rho_{0}$ is an overall normalization such that $\rho(r)$ is equal to the local density $\rho_{\odot} = 0.39$ GeV/cm$^3$ at $r =$ 8.5 kpc and $R_{s}=$ 24.42 kpc is a scale radius~\cite{Carlson:2014ssa}. In contrast to the spectra of \ahe\ from collisions of cosmic rays with the interstellar medium, the resulting spectrum for \ahe\ originating from DM annihilation peaks at low kinetic energies of around 0.1 GeV/$A$~\cite{Carlson:2014ssa}.

\subsection{Discussion of uncertainties on \ahe\ cosmic-ray modelling}
The results presented in this paper focus on the impact of the ALICE measurements for \sigmainel\ on the cosmic-ray \ahe\ flux and the corresponding transparency of the Galaxy. To this purpose, we have considered two models of \ahe\ sources described in the main text and only propagated the uncertainty of the \sigmainel\ measurement. Here we briefly discuss other possible uncertainties related to the \ahe\ cosmic-ray modelling.

As for the DM source, it is apparent that a different DM mass assumption changes the antinuclei flux profile near Earth~\cite{Ibarra:2012cc,Carlson:2014ssa,vonDoetinchem:2020vbj}. The DM mass assumptions around $m_{\chi} \sim 100$ GeV are favoured by recent AMS-02 antiproton data~\cite{vonDoetinchem:2020vbj}; for very different values of $m_{\chi}$, the \ahe\ flux and the corresponding transparency can be studied as described in this work. Variation of the DM annihilation cross section $\langle\sigma v\rangle$ leads to a constant scaling of \ahe\ flux according to Eq.~\ref{eqn:SourceFromDM} and therefore to identical transparency values. Although the Navarro–Frenk–White profile is used in this work to describe the distribution of DM in the Galaxy, other profiles are also available such as Einasto~\cite{Ibarra:2012cc}, Burkert~\cite{Burkert:1995yz} or the isothermal one~\cite{Begeman:1991iy}. 
Antiproton limits on $\langle\sigma v\rangle$ are partially degenerate with the effect of different DM profiles, and the overall impact of varying the DM profiles on the maximum allowed antinuclei flux is minor ~\cite{Korsmeier:2017xzj,Serksnyte:2022onw}.
If the isothermal profile is employed instead of the Navarro–Frenk–White one, the obtained \ahe\ transparency is shifted up by $10-15\%$.

Although the coalescence-based models can successfully describe antinuclei production, the model uncertainties are still relatively large, which leads to substantial changes in the magnitude of antinuclei fluxes~\cite{Serksnyte:2022onw,Korsmeier:2017xzj,Ibarra:2012cc}. In general, as long as different coalescence models retain the shape of the produced antinuclei momentum spectrum, the resulting transparency is not affected. For example, the change in coalescence parameter $p_{0}$ leads to constant scaling of the antinuclei flux and identical transparency values.

The GALPROP parameters used in this work are tuned to reproduce the available experimental data on cosmic-ray nuclei (up to $Z = 28$). The obtained uncertainties on the nuclei fluxes of $\lesssim 10\%$~\cite{Boschini:2020jty} are not considered in this work, since they result in a negligible change to the \ahe\ fluxes. An alternative set of propagation parameters has been obtained~\cite{Cuoco:2016eej} by considering a subsample of available cosmic-ray data. The comparison between the two sets is discussed in more details elsewhere~\cite{Serksnyte:2022onw}. The employment of these alternative parameters decreases the \ahe\ background flux by one order of magnitude at the lowest $E_{\rm{kin}}$ considered in this work and results in about 60\% lower transparency. For the DM signal, the corresponding flux is up to a factor of five higher at the lowest $E_{\rm{kin}}$ with about 40\% lower transparency. These differences in fluxes and transparencies are obtained before the solar modulation and become minor for $E_{\rm{kin}} \gtrsim 10$ GeV/$A$, for the DM signal as well as the background.

\section{Extended Data}
\label{sec:ExtendedData}

%%%%%%%%% appendix with author list
\newpage
\appendix
\section{The ALICE Collaboration}
\label{app:collab}
% ALICE Collaboration author list for 2021-12-14
\small
\begin{flushleft} 

S.~Acharya$^{\rm 142}$, 
D.~Adamov\'{a}$^{\rm 96}$, 
A.~Adler$^{\rm 74}$, 
J.~Adolfsson$^{\rm 81}$, 
G.~Aglieri Rinella$^{\rm 34}$, 
M.~Agnello$^{\rm 30}$, 
N.~Agrawal$^{\rm 54}$, 
Z.~Ahammed$^{\rm 142}$, 
S.~Ahmad$^{\rm 16}$, 
S.U.~Ahn$^{\rm 76}$
% ...

\section*{Affiliation Notes} 

$^{\rm I}$ Deceased\\
$^{\rm II}$ Also at: Italian National Agency for New Technologies, Energy and Sustainable Economic Development (ENEA), Bologna, Italy\\
$^{\rm III}$ Also at: Dipartimento DET del Politecnico di Torino, Turin, Italy\\
$^{\rm IV}$ Also at: M.V. Lomonosov Moscow State University, D.V. Skobeltsyn Institute of Nuclear, Physics, Moscow, Russia\\
$^{\rm V}$ Also at: Department of Applied Physics, Aligarh Muslim University, Aligarh, India\\
$^{\rm VI}$ Also at: Institute of Theoretical Physics, University of Wroclaw, Poland\\
$^{\rm VII}$ Also at: University of Kansas, Lawrence, Kansas, United States\\

\section*{Collaboration Institutes}

$^{1}$ A.I. Alikhanyan National Science Laboratory (Yerevan Physics Institute) Foundation, Yerevan, Armenia\\
$^{2}$ AGH University of Science and Technology, Cracow, Poland\\
$^{3}$ Bogolyubov Institute for Theoretical Physics, National Academy of Sciences of Ukraine, Kiev, Ukraine\\
$^{4}$ Bose Institute, Department of Physics  and Centre for Astroparticle Physics and Space Science (CAPSS), Kolkata, India\\
$^{5}$ Budker Institute for Nuclear Physics, Novosibirsk, Russia\\
$^{6}$ California Polytechnic State University, San Luis Obispo, California, United States\\
$^{7}$ Central China Normal University, Wuhan, China\\
$^{8}$ Centro de Aplicaciones Tecnol\'{o}gicas y Desarrollo Nuclear (CEADEN), Havana, Cuba\\
$^{9}$ Centro de Investigaci\'{o}n y de Estudios Avanzados (CINVESTAV), Mexico City and M\'{e}rida, Mexico\\
$^{10}$ Chicago State University, Chicago, Illinois, United States\\
$^{11}$ China Institute of Atomic Energy, Beijing, China\\
$^{12}$ Chungbuk National University, Cheongju, Republic of Korea\\
$^{13}$ Comenius University Bratislava, Faculty of Mathematics, Physics and Informatics, Bratislava, Slovakia\\
$^{14}$ COMSATS University Islamabad, Islamabad, Pakistan\\
$^{15}$ Creighton University, Omaha, Nebraska, United States\\
$^{16}$ Department of Physics, Aligarh Muslim University, Aligarh, India\\
$^{17}$ Department of Physics, Pusan National University, Pusan, Republic of Korea\\
$^{18}$ Department of Physics, Sejong University, Seoul, Republic of Korea\\
$^{19}$ Department of Physics, University of California, Berkeley, California, United States\\
$^{20}$ Department of Physics, University of Oslo, Oslo, Norway\\
$^{21}$ Department of Physics and Technology, University of Bergen, Bergen, Norway\\
$^{22}$ Dipartimento di Fisica dell'Universit\`{a} and Sezione INFN, Cagliari, Italy\\
$^{23}$ Dipartimento di Fisica dell'Universit\`{a} and Sezione INFN, Trieste, Italy\\
$^{24}$ Dipartimento di Fisica dell'Universit\`{a} and Sezione INFN, Turin, Italy\\
$^{25}$ Dipartimento di Fisica e Astronomia dell'Universit\`{a} and Sezione INFN, Bologna, Italy\\
$^{26}$ Dipartimento di Fisica e Astronomia dell'Universit\`{a} and Sezione INFN, Catania, Italy\\
$^{27}$ Dipartimento di Fisica e Astronomia dell'Universit\`{a} and Sezione INFN, Padova, Italy\\
$^{28}$ Dipartimento di Fisica e Nucleare e Teorica, Universit\`{a} di Pavia, Pavia, Italy\\
$^{29}$ Dipartimento di Fisica `E.R.~Caianiello' dell'Universit\`{a} and Gruppo Collegato INFN, Salerno, Italy\\
$^{30}$ Dipartimento DISAT del Politecnico and Sezione INFN, Turin, Italy\\
$^{31}$ Dipartimento di Scienze e Innovazione Tecnologica dell'Universit\`{a} del Piemonte Orientale and INFN Sezione di Torino, Alessandria, Italy\\
$^{32}$ Dipartimento di Scienze MIFT, Universit\`{a} di Messina, Messina, Italy\\
$^{33}$ Dipartimento Interateneo di Fisica `M.~Merlin' and Sezione INFN, Bari, Italy\\
$^{34}$ European Organization for Nuclear Research (CERN), Geneva, Switzerland\\
$^{35}$ Faculty of Electrical Engineering, Mechanical Engineering and Naval Architecture, University of Split, Split, Croatia\\
$^{36}$ Faculty of Engineering and Science, Western Norway University of Applied Sciences, Bergen, Norway\\
$^{37}$ Faculty of Nuclear Sciences and Physical Engineering, Czech Technical University in Prague, Prague, Czech Republic\\
$^{38}$ Faculty of Science, P.J.~\v{S}af\'{a}rik University, Ko\v{s}ice, Slovakia\\
$^{39}$ Frankfurt Institute for Advanced Studies, Johann Wolfgang Goethe-Universit\"{a}t Frankfurt, Frankfurt, Germany\\
$^{40}$ Fudan University, Shanghai, China\\
$^{41}$ Gangneung-Wonju National University, Gangneung, Republic of Korea\\
$^{42}$ Gauhati University, Department of Physics, Guwahati, India\\
$^{43}$ Helmholtz-Institut f\"{u}r Strahlen- und Kernphysik, Rheinische Friedrich-Wilhelms-Universit\"{a}t Bonn, Bonn, Germany\\
$^{44}$ Helsinki Institute of Physics (HIP), Helsinki, Finland\\
$^{45}$ High Energy Physics Group,  Universidad Aut\'{o}noma de Puebla, Puebla, Mexico\\
$^{46}$ Hiroshima University, Hiroshima, Japan\\
$^{47}$ Hochschule Worms, Zentrum  f\"{u}r Technologietransfer und Telekommunikation (ZTT), Worms, Germany\\
$^{48}$ Horia Hulubei National Institute of Physics and Nuclear Engineering, Bucharest, Romania\\
$^{49}$ Indian Institute of Technology Bombay (IIT), Mumbai, India\\
$^{50}$ Indian Institute of Technology Indore, Indore, India\\
$^{51}$ Indonesian Institute of Sciences, Jakarta, Indonesia\\
$^{52}$ INFN, Laboratori Nazionali di Frascati, Frascati, Italy\\
$^{53}$ INFN, Sezione di Bari, Bari, Italy\\
$^{54}$ INFN, Sezione di Bologna, Bologna, Italy\\
$^{55}$ INFN, Sezione di Cagliari, Cagliari, Italy\\
$^{56}$ INFN, Sezione di Catania, Catania, Italy\\
$^{57}$ INFN, Sezione di Padova, Padova, Italy\\
$^{58}$ INFN, Sezione di Pavia, Pavia, Italy\\
$^{59}$ INFN, Sezione di Torino, Turin, Italy\\
$^{60}$ INFN, Sezione di Trieste, Trieste, Italy\\
$^{61}$ Inha University, Incheon, Republic of Korea\\
$^{62}$ Institute for Gravitational and Subatomic Physics (GRASP), Utrecht University/Nikhef, Utrecht, Netherlands\\
$^{63}$ Institute for Nuclear Research, Academy of Sciences, Moscow, Russia\\
$^{64}$ Institute of Experimental Physics, Slovak Academy of Sciences, Ko\v{s}ice, Slovakia\\
$^{65}$ Institute of Physics, Homi Bhabha National Institute, Bhubaneswar, India\\
$^{66}$ Institute of Physics of the Czech Academy of Sciences, Prague, Czech Republic\\
$^{67}$ Institute of Space Science (ISS), Bucharest, Romania\\
$^{68}$ Institut f\"{u}r Kernphysik, Johann Wolfgang Goethe-Universit\"{a}t Frankfurt, Frankfurt, Germany\\
$^{69}$ Instituto de Ciencias Nucleares, Universidad Nacional Aut\'{o}noma de M\'{e}xico, Mexico City, Mexico\\
$^{70}$ Instituto de F\'{i}sica, Universidade Federal do Rio Grande do Sul (UFRGS), Porto Alegre, Brazil\\
$^{71}$ Instituto de F\'{\i}sica, Universidad Nacional Aut\'{o}noma de M\'{e}xico, Mexico City, Mexico\\
$^{72}$ iThemba LABS, National Research Foundation, Somerset West, South Africa\\
$^{73}$ Jeonbuk National University, Jeonju, Republic of Korea\\
$^{74}$ Johann-Wolfgang-Goethe Universit\"{a}t Frankfurt Institut f\"{u}r Informatik, Fachbereich Informatik und Mathematik, Frankfurt, Germany\\
$^{75}$ Joint Institute for Nuclear Research (JINR), Dubna, Russia\\
$^{76}$ Korea Institute of Science and Technology Information, Daejeon, Republic of Korea\\
$^{77}$ KTO Karatay University, Konya, Turkey\\
$^{78}$ Laboratoire de Physique des 2 Infinis, Ir\`{e}ne Joliot-Curie, Orsay, France\\
$^{79}$ Laboratoire de Physique Subatomique et de Cosmologie, Universit\'{e} Grenoble-Alpes, CNRS-IN2P3, Grenoble, France\\
$^{80}$ Lawrence Berkeley National Laboratory, Berkeley, California, United States\\
$^{81}$ Lund University Department of Physics, Division of Particle Physics, Lund, Sweden\\
$^{82}$ Moscow Institute for Physics and Technology, Moscow, Russia\\
$^{83}$ Nagasaki Institute of Applied Science, Nagasaki, Japan\\
$^{84}$ Nara Women{'}s University (NWU), Nara, Japan\\
$^{85}$ National and Kapodistrian University of Athens, School of Science, Department of Physics , Athens, Greece\\
$^{86}$ National Centre for Nuclear Research, Warsaw, Poland\\
$^{87}$ National Institute of Science Education and Research, Homi Bhabha National Institute, Jatni, India\\
$^{88}$ National Nuclear Research Center, Baku, Azerbaijan\\
$^{89}$ National Research Centre Kurchatov Institute, Moscow, Russia\\
$^{90}$ Niels Bohr Institute, University of Copenhagen, Copenhagen, Denmark\\
$^{91}$ Nikhef, National institute for subatomic physics, Amsterdam, Netherlands\\
$^{92}$ NRC Kurchatov Institute IHEP, Protvino, Russia\\
$^{93}$ NRC \guillemotleft Kurchatov\guillemotright  Institute - ITEP, Moscow, Russia\\
$^{94}$ NRNU Moscow Engineering Physics Institute, Moscow, Russia\\
$^{95}$ Nuclear Physics Group, STFC Daresbury Laboratory, Daresbury, United Kingdom\\
$^{96}$ Nuclear Physics Institute of the Czech Academy of Sciences, \v{R}e\v{z} u Prahy, Czech Republic\\
$^{97}$ Oak Ridge National Laboratory, Oak Ridge, Tennessee, United States\\
$^{98}$ Ohio State University, Columbus, Ohio, United States\\
$^{99}$ Petersburg Nuclear Physics Institute, Gatchina, Russia\\
$^{100}$ Physics department, Faculty of science, University of Zagreb, Zagreb, Croatia\\
$^{101}$ Physics Department, Panjab University, Chandigarh, India\\
$^{102}$ Physics Department, University of Jammu, Jammu, India\\
$^{103}$ Physics Department, University of Rajasthan, Jaipur, India\\
$^{104}$ Physikalisches Institut, Eberhard-Karls-Universit\"{a}t T\"{u}bingen, T\"{u}bingen, Germany\\
$^{105}$ Physikalisches Institut, Ruprecht-Karls-Universit\"{a}t Heidelberg, Heidelberg, Germany\\
$^{106}$ Physik Department, Technische Universit\"{a}t M\"{u}nchen, Munich, Germany\\
$^{107}$ Politecnico di Bari and Sezione INFN, Bari, Italy\\
$^{108}$ Research Division and ExtreMe Matter Institute EMMI, GSI Helmholtzzentrum f\"ur Schwerionenforschung GmbH, Darmstadt, Germany\\
$^{109}$ Russian Federal Nuclear Center (VNIIEF), Sarov, Russia\\
$^{110}$ Saha Institute of Nuclear Physics, Homi Bhabha National Institute, Kolkata, India\\
$^{111}$ School of Physics and Astronomy, University of Birmingham, Birmingham, United Kingdom\\
$^{112}$ Secci\'{o}n F\'{\i}sica, Departamento de Ciencias, Pontificia Universidad Cat\'{o}lica del Per\'{u}, Lima, Peru\\
$^{113}$ St. Petersburg State University, St. Petersburg, Russia\\
$^{114}$ Stefan Meyer Institut f\"{u}r Subatomare Physik (SMI), Vienna, Austria\\
$^{115}$ SUBATECH, IMT Atlantique, Universit\'{e} de Nantes, CNRS-IN2P3, Nantes, France\\
$^{116}$ Suranaree University of Technology, Nakhon Ratchasima, Thailand\\
$^{117}$ Technical University of Ko\v{s}ice, Ko\v{s}ice, Slovakia\\
$^{118}$ The Henryk Niewodniczanski Institute of Nuclear Physics, Polish Academy of Sciences, Cracow, Poland\\
$^{119}$ The University of Texas at Austin, Austin, Texas, United States\\
$^{120}$ Universidad Aut\'{o}noma de Sinaloa, Culiac\'{a}n, Mexico\\
$^{121}$ Universidade de S\~{a}o Paulo (USP), S\~{a}o Paulo, Brazil\\
$^{122}$ Universidade Estadual de Campinas (UNICAMP), Campinas, Brazil\\
$^{123}$ Universidade Federal do ABC, Santo Andre, Brazil\\
$^{124}$ University of Cape Town, Cape Town, South Africa\\
$^{125}$ University of Houston, Houston, Texas, United States\\
$^{126}$ University of Jyv\"{a}skyl\"{a}, Jyv\"{a}skyl\"{a}, Finland\\
$^{127}$ University of Kansas, Lawrence, Kansas, United States\\
$^{128}$ University of Liverpool, Liverpool, United Kingdom\\
$^{129}$ University of Science and Technology of China, Hefei, China\\
$^{130}$ University of South-Eastern Norway, Tonsberg, Norway\\
$^{131}$ University of Tennessee, Knoxville, Tennessee, United States\\
$^{132}$ University of the Witwatersrand, Johannesburg, South Africa\\
$^{133}$ University of Tokyo, Tokyo, Japan\\
$^{134}$ University of Tsukuba, Tsukuba, Japan\\
$^{135}$ University Politehnica of Bucharest, Bucharest, Romania\\
$^{136}$ Universit\'{e} Clermont Auvergne, CNRS/IN2P3, LPC, Clermont-Ferrand, France\\
$^{137}$ Universit\'{e} de Lyon, CNRS/IN2P3, Institut de Physique des 2 Infinis de Lyon, Lyon, France\\
$^{138}$ Universit\'{e} de Strasbourg, CNRS, IPHC UMR 7178, F-67000 Strasbourg, France, Strasbourg, France\\
$^{139}$ Universit\'{e} Paris-Saclay Centre d'Etudes de Saclay (CEA), IRFU, D\'{e}partment de Physique Nucl\'{e}aire (DPhN), Saclay, France\\
$^{140}$ Universit\`{a} degli Studi di Foggia, Foggia, Italy\\
$^{141}$ Universit\`{a} di Brescia, Brescia, Italy\\
$^{142}$ Variable Energy Cyclotron Centre, Homi Bhabha National Institute, Kolkata, India\\
$^{143}$ Warsaw University of Technology, Warsaw, Poland\\
$^{144}$ Wayne State University, Detroit, Michigan, United States\\
$^{145}$ Westf\"{a}lische Wilhelms-Universit\"{a}t M\"{u}nster, Institut f\"{u}r Kernphysik, M\"{u}nster, Germany\\
$^{146}$ Wigner Research Centre for Physics, Budapest, Hungary\\
$^{147}$ Yale University, New Haven, Connecticut, United States\\
$^{148}$ Yonsei University, Seoul, Republic of Korea\\

\end{flushleft} 
  %%%%%%% done by webmaster team
% TC:endignore
\end{document}